\documentclass{aa}
\usepackage{graphicx}
\usepackage{txfonts}

\begin{document}
   
   \title{Correlation between the TeV and X--ray emission in 
          high--energy peaked BL Lac objects.}
      
   \author{Katarzy\'nski K. \inst{1,2}, 
           Ghisellini G.,   \inst{1}, 
	   Tavecchio F.     \inst{1}
           Maraschi L.      \inst{1}\\
           Fossati G.       \inst{3}
           Mastichiadis A.  \inst{4}
	  }

   \offprints{Krzysztof Katarzy\'nski \\kat@astro.uni.torun.pl}

   \institute{Osservatorio Astronomico di Brera, via Bianchi 46, Merate and via Brera 28, 
              Milano, Italy  \and 
              Toru\'n Centre for Astronomy, Nicolaus Copernicus University, 
              ul. Gagarina 11, PL-87100 Toru\'n, Poland \and
              Rice University, Dept. of Physics and Astronomy, MS 108
                      6100 Main street, Houston, U.S.A.
              \and Dept. of Physics, Univ. of Athens, Panepistimiopolis, 15784 Zografos, Athens 
             }
	     
   \date{Received 29 June 2004 / 30 November 2004}

   \abstract{We discuss the correlation between the evolution of the TeV
             emission and X--ray radiation observed in high-energy peaked 
             BL Lac objects. We describe such a correlation by a simple 
             power law $F_{\rm TeV}(t) \propto F^x_{\rm X-ray}(t)$. 
             In the first part of this work we present correlations
             obtained for the activity of Mrk~501 observed in 1997 April and for
             the activity of Mrk~421 observed in 2000 February. Our results obtained 
             for Mrk~501 show that the index of the correlation ($x$) may strongly 
             depend on the width and position of the spectral bands used for the 
             comparison. The result of the correlation which we have obtained for 
             Mrk~421 is not informative. However, we discuss results of similar 
             correlation obtained for this source by other authors. They report an
             almost quadratic ($ x \sim 2$) correlations observed between the evolution 
             of the TeV and X--ray emission.
             In the second part of this paper we present a phenomenological model which 
             describes the evolution of the synchrotron and inverse Compton emission of 
             a simple spherical homogeneous source. 
             Neglecting the radiative cooling of the particles we derive analytical expressions 
             that describe the evolution. Then we use a numerical code to investigate the 
             impact of radiative cooling on the evolution. We show that different forms 
             of correlations can be obtained depending on the assumed evolution 
             scenario and the spectral bands used for the calculation. However,
             the quadratic correlation observed during the decay phase of
             the flare observed in Mrk~421 on 2001 March 19 appears
             problematic for this basic modeling. The quadratic
             correlation can be explained only for specific choices of the
             spectral bands used for the calculation. Therefore, looking
             for more robust solutions, we investigate the evolution of
             the emission generated by a cylindrical source.  However
             this model does not provide robust solutions for the
             problem of a quadratic correlation. In principle the problem
             could be solved by the TeV emission generated by the self
             Compton scattering in the Thomson limit.  However, we show
             that such a process requires unacceptably large values of the
             Doppler factor. Finally we briefly discuss the possible 
             influence of the light travel time effect on our results.
             }

\titlerunning{Correlation between the TeV and X--ray emission...}
\authorrunning{Katarzy\'nski et al.}   
\maketitle

\section{Introduction}

The emission from BL Lacertae objects is dominated by the intense 
relativistic boosted non--thermal continuum produced within a 
relativistic jet closely aligned with the line of sight, making 
these objects (together with Flat Spectrum Radio Quasars, the other 
subgroup of the {\it blazars} family) the best laboratories to study the 
physics of relativistic jets.

The overall emission, from radio to $\gamma$--rays (extended in some
cases to the multi--TeV band), shows the presence of two well--defined
broad components, the first one peaking in the optical--soft-X--ray
bands, the second one in the GeV--TeV region. The low energy peak is
attributed to synchrotron emission by relativistic
electrons in the jet, while the second component is commonly believed
to be Inverse Compton emission (hereafter IC) from the same electron 
population (although different scenarios have been proposed, see e.g.
Mannheim \cite{mannheim93}, Aharonian \cite{aharonian00}, Pohl \& 
Schlickeiser \cite{pohl00}, Mucke et al. \cite{mucke03}).

BL Lacs are further subdivided in Low--energy Peak BL Lacs (LBL),
exhibiting the synchrotron peak in the IR-optical region of the
spectrum, and High--energy Peak BL Lacs (HBL), in which the
synchrotron peak can lie in the UV--X--ray band. The almost
featureless optical spectra of HBL clearly indicate that the
environment external to the jet is quite poor in soft photons,
suggesting that the high--energy emission is mainly produced through
the Comptonization of the synchrotron radiation (Synchrotron
Self--Compton emission, SSC). The few extragalactic TeV sources firmly
detected so far (Mrk~421, Mrk 501, PKS 2344+514, PKS 2155--304,
1ES1959+65, 1ES 1426+428, e.g. Krawczynski \cite{krawczynski04} for 
a recent review) belong to this class.

Since the discovery of the first BL Lac object emitting TeV radiation,
Mrk~421, (Punch et al. \cite{punch92}) TeV blazars have been the target of
very intense observational and theoretical investigations.  Indeed the
possibility of observing the emission produced by very high energy
electrons (up to Lorentz factors of the order of $10^7$) coupled with
observations in the X--ray band, where the synchrotron peak of these
sources is usually located, offers a unique tool to probe the
processes responsible for the cooling and the acceleration of
relativistic particles. 

Studies conducted simultaneously in the X--ray and in the TeV bands
are of particular importance, since in the simple SSC framework one
expects that variations in X--rays and TeV should be closely
correlated, being produced by electrons with similar energies.
Assuming a typical magnetic field intensity $B\sim 0.1$ G and Doppler
factor $\delta\sim 10$, photons with energy E $\sim 1$ keV are emitted
by electrons with Lorentz factor $\gamma \sim 10^6$. The same
electrons will upscatter photons at energy $E\sim \gamma mc^2\sim 1$
TeV (since the Lorentz factor is extremely large the scattering will
occur in the Klein--Nishina regime even with optical target photons).
In fact observations at X--ray and TeV energies (Buckley et al. \cite{buckley96},
Catanese et al. \cite{catanese97}) yielded significant evidence of correlated and
simultaneous variability of the TeV and X--ray fluxes.  During the
X/TeV 1998 campaign on Mrk~421 a flare was detected
simultaneously both at X--ray and TeV energies and the maxima were
simultaneous within 1 hour, confirming that variations in these two
bands are closely related (Maraschi et al. \cite{maraschi99}). Subsequent 
more extensive analyses confirmed these first results also in other
sources. Note however that the correlation seems to be violated in
some cases, as indicated by the observation of an ``orphan'' (i.e. not
accompanied by the corresponding X--ray flare) TeV event in the BL Lac
1ES 1959+650 (Krawczynski et al. \cite{krawczynski04a}).

Although the study of TeV BL Lacs
has been the subject of great effort in the past (e.g. 
B\"ottcher \cite{bottcher04}, Georganopoulos \& Kazanas \cite{georg03},
Moderski, Sikora \& Blazejowski \cite{moderski03}, B\"ottcher \&
Dermer \cite{bottcher02b}, Tavecchio et al. \cite{tavecchio01},
Takahashi et al. \cite{takahashi00}, Kirk, Rieger \& Mastichiadis \cite{kirk98}),
a detailed analysis of the expected correlations of the fluxes in different 
bands is lacking. In this paper we present for the first time a
comprehensive study of the time--dependent emission expected in the
case of the homogeneous SSC model, following the evolution of
electrons and taking into account both radiative and adiabatic
losses. In particular we focus our attention on the correlation between
the variations observed in the X--rays and in the TeV band, which can
be quite informative of the processes responsible for the observed
variations. The present work has been stimulated (as described in
Sect. 2) by the observation in several cases of flares for which a
more than linear behavior was observed 
(namely, $F_{\rm TeV} \propto F^x_{\rm X-ray}$, with $x>1$).
At first sight this does not pose particular
problems (an increase of the density of the emitting electrons
will make the self--Compton flux increase more than the synchrotron
flux), but we will show instead that these more than linear
correlations are quite difficult to reproduce,
with standard assumptions, when we consider that the
same type of correlation has been observed not only
during the rising phases of a flare, but also during the decay phase
(Fossati at al. \cite{fossati04}).

In Section 2 we discuss some of the clearest observational evidence for
the correlation between X--rays and $\gamma$--rays. In Section 3 and 4 we
describe the model used to calculate the expected correlations in the SSC
framework (assuming spherical and cylindrical geometries) and we present
the results, stressing the difficulty of obtaining more than linear relations
between TeV and X-ray variations. In Section 5 we briefly discuss the effects 
that can be induced by considering the finite light crossing time in the 
study of the variability. In Section 6 we conclude. 

\section{Observed correlations}
\label{sec_obs-corr}

The correlation between the evolution of the X--ray and the $\gamma$--ray
emission of high--energy peaked BL Lac (HBL) objects may provide important
information on the emission mechanism of such sources and the physical
mechanisms producing the variability. However, to investigate such a
correlation it is necessary to compare two different light curves,
obtained in the same period of time by two different instruments. The
sampling rate for both light curves must be similar to provide a
reliable result.  Since the variability timescale in HBL objects is very
short, the sampling of the light curves should also be dense, with
several data points per day.  Moreover, additional simultaneous spectral
observations are required to provide basic information about the
evolution of the spectrum.  The last condition is very important for the
modeling of the source activity. For example it is important to know if
the X--ray synchrotron emission ($F_s$) is generated below or above the
$\nu F_{s}(\nu)$ peak. It is difficult to meet all these conditions.
Therefore, there are just a few cases where the correlation between the
X--ray and the $\gamma$--ray activity can be investigated in detail.
In this section we present three correlations obtained for two very well
known HBL objects Mrk~421 and Mrk~501, and we also discuss the
correlations obtained by other investigators.

\phantom{nothing}
\begin{figure}[p]
\resizebox{\hsize}{!}{\includegraphics{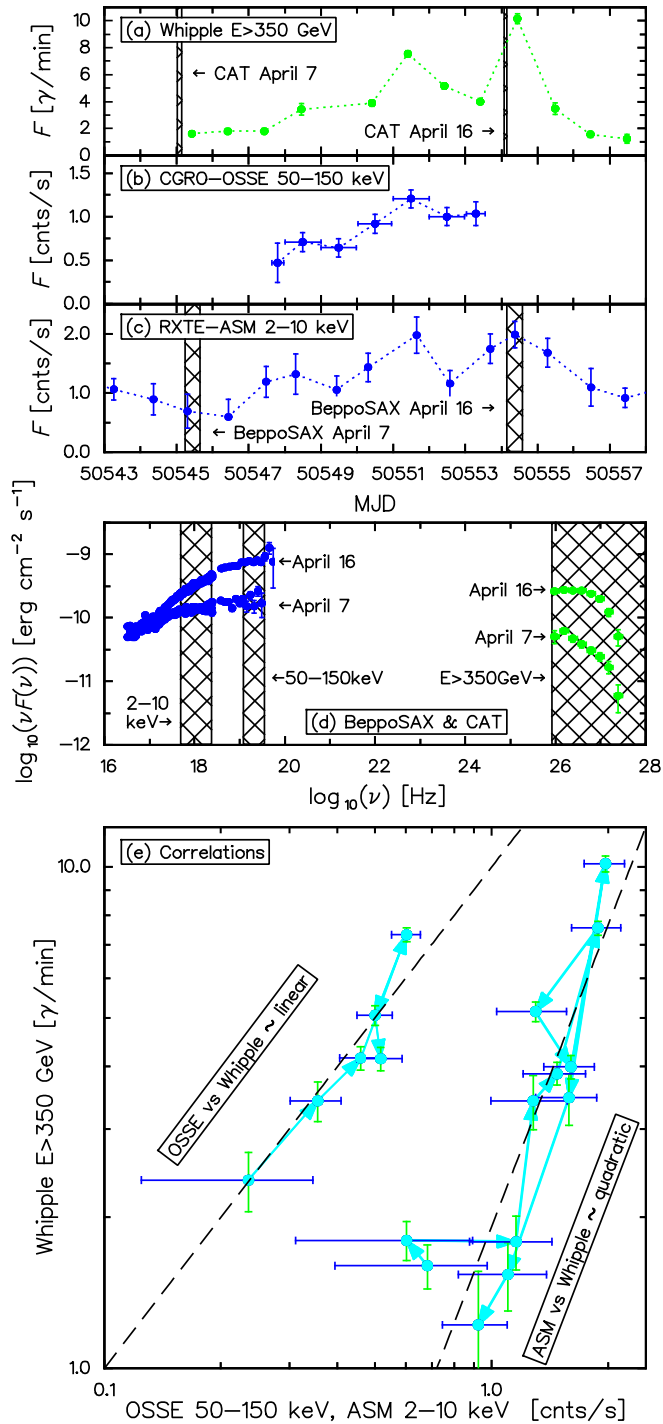}}
\caption{The activity of Mrk~501 observed in 1997 April. The upper panels show 
         the $\gamma$--ray and the X--ray light curves obtained by the Whipple (a), CGRO--OSSE (b)
         and RXTE--ASM (c) experiments (Catanese et al. \cite{catanese97}). The shaded areas in
         these figures indicate the observing periods of the $Beppo$SAX and CAT instruments.
         In panel (d) we show the spectra obtained by the above mentioned instruments (Pian
         et al. \cite{pian98}, Djannati--Atai et al. \cite{djannati99}). In this panel we 
         show also the energy bands used to obtain the light curves. The lower panel (e) 
         shows the two correlations made between the X--ray and the TeV gamma--ray fluxes.
         For clarity, we multiply the OSSE data in this panel by the factor 0.5. 
         The dashed lines in this panel show a template for the linear and the quadratic 
         correlation.
        }
\label{fig_501corr}        
\end{figure}

\begin{figure}[!t]
\resizebox{\hsize}{!}{\includegraphics{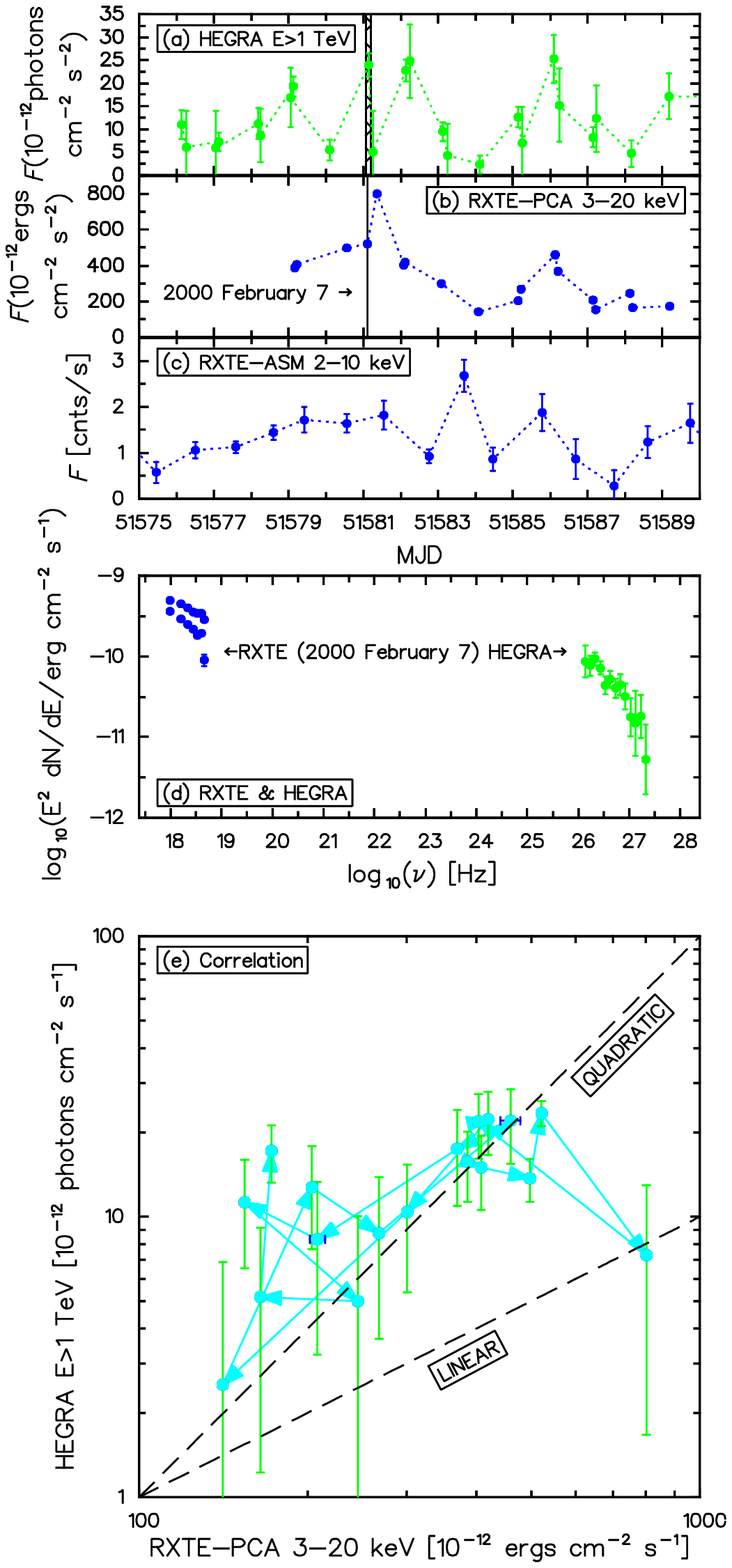}}
\caption{The activity of Mrk~421 observed on 2000 February. The upper panels show 
         the $\gamma$--ray and the X--ray light curves obtained by the HEGRA (a), RXTE--PCA (b)
         and RXTE--ASM (c) experiments (Krawczynski et al. \cite{krawczynski01}). The shaded 
         areas in these panels indicate the observing periods for the spectra obtained by the 
         RXTE and HEGRA instruments presented in panel (d). The lower panel (e) shows the 
         correlation between the X--ray and the TeV $\gamma$--ray fluxes. The dashed lines 
         in this panel show a template for the linear and the quadratic correlation.
~~~~~~~~~~~~~~~~~~~~~~~~~~~~~~~~~~~~~~~~~~~~~~~~~~~~~~~~~~~~~~~~~~~~~~~~~~~~~~~~~~~~~~~~~~~~~~~~~~~~~~~
~~~~~~~~~~~~~~~~~~~~~~~~~~~~~~~~~~~~~~~~~~~~~~~~~~~~~~~~~~
~~~~~~~~~~~~~~~~~~~~~~~~~~~~~~~~~~~~~~~~~~~~~~~~~~~~~~~~~~~~~~~~~~~~~~~~~~~~~~~~~~~
        }
\label{fig_421corr}        
\end{figure}

In Figure \ref{fig_501corr} we show the activity of Mrk~501 observed in 1997 April. We show
in this figure the light curves obtained by the Whipple (E$>$350 GeV), CGRO--OSSE (50--150 keV), 
and RXTE--ASM (3--20 keV) experiments (Figs \ref{fig_501corr}--a,b,c respectively, Catanese et al. 
\cite{catanese97}). We show also the evolution of the spectral energy distribution (Fig. 
\ref{fig_501corr}--d) observed by the $Beppo$SAX (0.1--200 keV) and CAT (E$>$250 GeV) instruments 
during this activity (Pian et al. \cite{pian98}, Djannati--Atai et al. \cite{djannati99}). For 
this data we calculate two correlations. The first correlation is calculated between the 
fluxes obtained by the OSSE and the Whipple experiments. This correlation gives an almost 
linear result (Fig. \ref{fig_501corr}--e). The precise computation gives in this case 
$F_{\rm{Whipple}} \propto F_{\rm{OSSE}}^{1.71\pm0.50}$ with a reduced $\chi^2 = 0.67$. 
The second correlation is calculated between the data from the ASM experiment and the fluxes 
obtained by the Whipple telescope. This correlation gives a quadratic or even more than quadratic 
result. The precise computation gives $F_{\rm{Whipple}} \propto 
F_{\rm{ASM}}^{2.69\pm0.56}$ and the reduced $\chi^2 = 0.65$. 
Note that to compare two light curves, where the observational points are randomly
distributed in time, it is necessary to select points gathered almost exactly at the same 
time. This may practically mean that we are able to correlate only a few points in a light
curves where we have even dozens of points. In the alternate approach we may try to
interpolate one of the light curves. However, this method requires similar sampling of
the light curves. Moreover, the sampling should be comparable or less than the characteristic 
variability time scale of a source. In the case of the OSSE-Whipple correlation we have 
interpolated the Whipple light curve. The interpolation method in this particular case 
has a minimal influence on the final result because the OSSE and Whipple observational 
points were gathered almost at the same time. There is only one point in the 
OSSE light curve (50549-50550 MJD) which has no counterpart in the TeV observations 
and we have excluded this point from our analysis. In the 
ASM-Whipple correlation we have interpolated the ASM light curve. The time shifts between 
the observational points in these light curves are relatively large in comparison to the 
previous correlation. However, each point in the ASM light
curve has been calculated as an average of many basic measurements made during a period
of about one day. Therefore, the interpolation in this case also provides meaningful 
results. Moreover, our results are in agreement with the 
previous correlation made by Djannati--Atai et al. (\cite{djannati99}). 
They obtained a more than linear but less than quadratic correlation between the 
$Beppo$SAX and the CAT telescope fluxes. 
Note that the energy range of the $Beppo$SAX observations (0.1--200 keV)
covers the spectral bands used by the ASM and the OSSE experiments 
(Fig. \ref{fig_501corr}--d).
Therefore, the correlation mentioned above gives an average of the two correlations presented in
this work. 
The source was observed at that time also by other instruments. Krawczynski et al.
(\cite{krawczynski00}) correlated the evolution of the X--ray emission at 3 keV with the 
evolution of the $\gamma$--ray radiation at 2 TeV. The correlation they obtained seems to
be more than quadratic (see Figure 6--a in their paper). 
This is very similar to the  
correlation presented here
between the ASM (2--10 keV) and the Whipple observations. The second correlation 
performed by the team mentioned above concerns the evolution of the X--ray emission at 25 keV 
and the evolution of the $\gamma$--rays at 2 TeV. 
This correlation gives an almost quadratic result
(see Figure 6--b in their article).

The correlations discussed above show that there was significant 
difference between the evolution of the synchrotron emission before 
the $\nu F_s(\nu)$ peak and the evolution around the peak. It means 
that the correlation may depend on the position of the spectral bands 
(below, around or above the $\nu F_{s/c}(\nu)$ peak) used for the 
calculations. The comparison between our correlations and the correlation 
done by Djannati--Atai et al. (\cite{djannati99}) indicates that the 
slope of the correlation may also depend on the width of the spectral band.

In Figure \ref{fig_421corr} we present the activity of Mrk~421 observed 
in 2000 February. We show the light curves gathered by the HEGRA (E$>$1TeV), 
RXTE--PCA (3--20 keV) and the RXTE--ASM (2--10 keV) experiments 
(Figs \ref{fig_501corr}--a,b,c respectively, Krawczynski et al. \cite{krawczynski01}). 
We show also the spectral energy distributions observed at that time 
(Fig. \ref{fig_501corr}--d). We calculate for this source the correlation 
between the PCA and the HEGRA fluxes. To obtain the correlation we have 
interpolated the HEGRA light curve. However, the interpolation as 
in the case of the OSSE-Whipple correlation for Mrk 501, has minimal 
influence on the result because the observations in both light 
curves were made almost simultaneously. In this case we have obtain 
unclear results; the detailed fitting gives $F_{\rm{HEGRA}} \propto 
F_{\rm{PCA}}^{0.47\pm0.15}$ with a poor reduced $\chi^2$ 
of 1.7. Note that also the correlation between the ASM and the HEGRA data, 
which we do not present here, does not provide an informative result. 
This example shows that the relatively long timescale observations with 
one data point per day may not be sufficient to obtain meaningful 
information on the TeV/X--ray correlation.

However, Fossati et al. (\cite{fossati04}) have recently reported very precise observations 
of this source made in 2001 March. The object was observed during a period of one 
week by the RXTE--PCA, Whipple and the HEGRA instruments. The rate of the measurements 
was relatively high and several flaring events were well observed. The most significant 
seems to be a flare observed on March 19 by the PCA and Whipple instruments. 
The TeV/X--ray correlation for this flare is almost quadratic ($F_{\rm TeV} \propto 
F_{\rm X}^{2.3\pm0.3}$) for the rising and decaying phase. However, the correlation 
for data obtained by the Whipple and RXTE--PCA experiments during this campaign
gives $F_{\rm TeV} \propto F_{\rm X}^{1.3\pm0.1}$. This indicates that the longterm 
correlation may give results significantly different from the correlation performed 
for one single flaring event. The reason for the difference may be related to the fact 
that for a long term light curve (e.g. a few days or even longer) we may correlate at 
least a few flaring events. If for example the activity is generated in the framework 
of the internal shock mechanism (e.g. Rees \cite{rees78}, Guetta et al. \cite{guetta04}) 
then the flares may overlap each other in time. If we observe a decay phase of a flare 
and during this decay a new flare will start to rise then the observed correlation
would be completely different even if the base correlation for each flare was the same.

Recently Tanihata et al. (\cite{tanihata04}) published a spectral analysis of 
the observations of Mrk~421 obtained during a 7 day campaign in 1998. They found 
a correlation between the TeV flux and the synchrotron peak flux $F_{\rm TeV} \propto 
F_{\rm s,peak}^{1.7\pm0.3}$. 

In the following we will take particular care in looking for robust
solutions yielding linear or quadratic correlations between X-ray and TeV
bands. Although so far there are only few cases for which these
correlations (especially the quadratic one) have been clearly
established, our attention is motivated by the fact that, as we will
discuss below, the observations of such a correlations could impose
strong constraints on the scenario usually considered for the variability
observed in these sources. In particular the observation of a quadratic
relation during the {\it decaying phase} is particularly intriguing. In
fact, if the quadratic increase of the synchrotron and IC emission could
be easily reproduced, increasing the electron density within the source,
the decrease is more problematic. Escape of electrons from the
sources appears unrealistic in the small timescale shown by
variability. On the other hand radiative cooling can affect only the
high-energy particles. Indeed, the simple quadratic relation predicted
for the IC emission is strictly valid for IC in the Thomson regime. 
However, the TeV emission of the HBL objects is probably generated in the
Klein-Nishina regime. In Appendix \ref{app_thomson} we estimate the physical 
parameters of a source that could generate TeV emission in
the Thomson limit. The estimation gives unacceptably large values of the
Doppler factor ($\delta \gtrsim 1000$, Begelman et al. \cite{begelman94}). 
Therefore, in all models presented in this work we assume that the TeV radiation 
comes from the IC scattering in the  Klein-Nishina regime.
The decline of the cross-section in the Klein-Nishina regime (valid for
sources with a plausible value of the physical quantities) has the effect
of decoupling the (high-energy) electrons producing the high-energy photons
and those (at low-energy) producing the optical--IR synchrotron seed
photons (e.g. Tavecchio, Maraschi \& Ghisellini 1998). In these
conditions, since the radiative cooling only affects the high-energy
particles, a {\it linear} relation between X-rays and TeV should be
expected, since the seed photons can be considered almost constant during
the decay. A plausible alternative, discussed in detail below, is to
admit adiabatic cooling, affecting all the electrons, irrespective of
their energy.

\section{Spherical homogeneous source}
\label{sec_spher-src}

A homogeneous synchrotron-SSC model is frequently proposed as a possible
explanation for the X--ray and the $\gamma$--ray emission of HBL
objects. Such model provide a very good and simple explanation for
the Spectral Energy Distributions (SED) observed in X--rays and the
TeV gamma rays (e.g. Dermer $\&$ Schlickeiser \cite{dermer93},
Bednarek $\&$ Protheroe \cite{bednarek97}, Mastichiadis 
\& Kirk \cite{mastichiadis97}, Pian et al. \cite{pian98}, Katarzy\'nski et al. 
\cite{katarzynski01}). Therefore, we decided to check if such a model is
also able to explain quite specific properties of the observed
variability.  In the approach presented in this section we do not
consider Light Crossing Time Effects (hereafter LCTE) in the synchrotron
radiation field inside the source nor for the total observed emission.
This means that the results presented in this section are strictly
valid only if the physical processes, which modify the source brightness,
are slower than the source light crossing time.  This condition may
not always be correct in the case of blazars. However, it allows us to
investigate in detail the most important physical processes which may
modify the emission level of the source. Detailed understanding of these
processes is very important for any more complex modeling.

\subsection{Description of the model}
\label{sub_sph-geom} 

We assume for this modeling a spherical homogeneous source, which may
undergo expansion or compression. As a first step, to simplify the model and
to provide analytical formulae which describe the evolution of such a
source, we decided to consider only four physical processes which may
occur during the evolution. We calculate increase or decrease of the
source volume, increase or decay of the magnetic field intensity,
variations of the particle density and adiabatic heating or cooling of
the particles. In the model we separate as much as possible the
mathematical description for each process. This allows us to
investigate separately the influence of each process on the source
emission. In this first approach we do not consider the radiative cooling
of the particles. The impact of this process will be discussed in the next
part of this work (Subsection \ref{sub_rad-cool}).

The time evolution of the source radius ($R$) and the magnetic field 
intensity ($B$) are assumed to be a power law functions
\begin{equation}
\label{equ_evol_magn}
R(t) = R_0 \left(t_0/t \right)^{-r_e},~~~~~
B(t) = B_0 \left(t_0/t \right)^{m},
\end{equation}
where $t_0$, $R_0$ and $B_0$ are initial time, magnetic field intensity and 
radius respectively. The indices $r_e$ and $m$ are a free parameters in 
our modeling.

The initial ($t= t_0$) distribution of the electron energy inside the source is defined 
by a broken power law
\begin{equation}
\label{equ_ini_elec_spec}
N_0(\gamma) = 
\left\{
\begin{array}{ll}
k_1 \gamma^{-n_1}, & \mbox{$ \gamma_{\rm min} \leq     \gamma \leq \gamma^0_{\rm brk} $}\\
k_2 \gamma^{-n_2}, & \mbox{$ \gamma^0_{\rm brk} ~<     \gamma \leq \gamma_{\rm max} $}\\
\end{array}
\right.,
\end{equation}
where $k_2 = k_1 (\gamma^0_{\rm brk})^{n_2-n_1}$, $\gamma$ is the Lorentz factor which is 
equivalent to the electron energy, $\gamma^0_{\rm brk}$ describes initial position of the 
break, $n_1$ and $n_2$ are spectral indices before and above the break respectively.
In the model we assume that the dominant part of the X-ray and TeV emission is produced by
the electrons with the Lorentz factors around $\gamma_{\rm brk}$. Therefore, the values of 
$\gamma_{\rm min}$ and $\gamma_{\rm max}$ parameters are not important for the results of the 
modeling. For all calculations presented in this section we use $\gamma_{\rm min} = 1$ and
$\gamma_{\rm max} = 10^8$.

The evolution of the electron energy spectrum is defined by a minimum
\begin{equation}
\label{equ_min_elec_spec}
N(\gamma, t) = \min\left\{N_1(\gamma, t), N_2(\gamma, t)\right\}
\end{equation}
of two power law functions
\begin{equation}
\label{equ_sph_pow_elec_spec}
\begin{array}{ll}
\vspace*{0.2cm}
N_1(\gamma, t) = K_1(t) \gamma^{-n_1}; &~~~~~K_1(t) = k_1 \left(\frac{t_0}{t} \right)^{3r_{d}}
                                                                \left(\frac{t_0}{t} \right)^{r_{a}(n_1-1)},\\
N_2(\gamma, t) = K_2(t) \gamma^{-n_2}; &~~~~~K_2(t) = k_2 \left(\frac{t_0}{t} \right)^{3r_{d}}
                                                                \left(\frac{t_0}{t} \right)^{r_{a}(n_2-1)},\\
\end{array}                                               
\end{equation}
where $K_1$ and $K_2$ describe the evolution of the particle density before and after the break 
($\gamma_{brk}$). The exponent $3r_d$ describes the decrease or increase of the particle density. 
The adiabatic heating or cooling of the particles is described by the indices $r_{a}(n_{1|2}-1)$. 
Note that if we assume for example adiabatic expansion or compression of the source with a constant 
number of the particles, then the parameters $r_e$, $r_d$ and $r_a$ should be equal (e.g. Kardashev 
\cite{kardashev62}, Longair \cite{longair92}). This is in fact the most realistic case. However, to 
investigate separately the influence of the above mentioned processes, we decided to produce 
such a detailed parameterization.

Assuming that the electron spectrum is defined by the minimum of two evolving power law functions we 
can easily find the evolution of the break $\gamma_{\rm brk}(t) = \gamma^0_{\rm brk} 
\left(t_0/t \right)^{r_{a}}.$

\subsubsection{Synchrotron emission}

The synchrotron emission coefficient which describes the radiation of the electrons is given by
the well-known formula $j_s \propto K B^{(\alpha+1)}$ (e.g. Rybicki $\&$ Lightman
\cite{rybicki79}) where $\alpha = (n-1)/2$. In our particular case we have to define 
the evolution of two emission coefficients
\begin{equation}
j^{1|2}_s(t) \propto K_{1|2}(t) B(t)^{(\alpha_{1|2}+1)} 
             \propto k_{1|2} B_{1|2} \left(\frac{t_0}{t}\right)^{3r_d+r_a(n_{1|2}-1)+m(\alpha_{1|2}+1)}, 
\end{equation}
where $j^1_s$ describes emission of the low energy electrons ($\gamma < \gamma_{brk}$), 
$j^2_s$ describes the radiation of the high energy electrons ($\gamma > \gamma_{brk}$) and
$B_{1|2} = B_0^{\alpha_{1|2}+1}$.\footnote{Note that in order to reduce the number of equations 
                                    we use in this work the following simplification 
                                    $f^{1|2}_x \propto a_{1|2}$ means $f^1_x \propto a_1$ 
                                    or $f^2_x \propto a_2$.}       

In our modeling we neglect the electron self--absorption process which is important only for the 
emission at the radio frequencies which we do not analyze in this work. Therefore, we can 
approximate the evolution of the intensity of the synchrotron radiation from the spherical 
source by
\begin{equation}
\label{equ_sph_isyn12}
I^{1|2}_s(t) \propto R(t) j^{1|2}_s(t) 
             \propto R_0 k_{1|2} B_{1|2} 
             \left(\frac{t_0}{t}\right)^{-r_e + 3r_d+r_a(n_{1|2}-1)+m(\alpha_{1|2}+1)}.
\end{equation}
Finally we can write the evolution of the synchrotron flux multiplying the intensity by the 
source surface
\begin{eqnarray}
\vspace*{0.1cm}
\label{equ_sph_fsyn1}
F^1_s(t) & \propto & R^2(t) I^1_s(t) ~\propto~ R_0^3 k_1 B_1 \left(\frac{t_0}{t}\right)^{-s_1},\\ 
    s_1  & =       & 3r_e - 3r_d - r_a(n_1-1) - m(\alpha_1+1),
\end{eqnarray}
\begin{eqnarray}                            
\vspace*{0.1cm}                            
\label{equ_sph_fsyn2}
F^2_s(t) & \propto & R^2(t) I^2_s(t) ~\propto~  R_0^3 k_2 B_2 \left(\frac{t_0}{t}\right)^{-s_2},\\
    s_2  & =       & 3r_e - 3r_d - r_a(n_2-1) - m(\alpha_2+1),
\end{eqnarray}
where the indices $s_1$ and $s_2$ describe the evolution of the synchrotron emission generated 
by the low and  the high energy electrons respectively. Note that there is a difference 
between the evolution of the $F^1_s$ flux and the evolution of the $F^2_s$ radiation due to the 
adiabatic processes ($r_a(n_{1|2}-1)$) and the evolution of the magnetic field intensity 
($m(\alpha_{1|2}+1)$).

\subsubsection{Self Compton emission}

Tavecchio et al. (\cite{tavecchio98}) presented detailed studies of the SSC emission generated 
by the electrons with an energy spectrum approximated by a broken power law. The IC spectrum in 
such an approach can be divided into four basic components ($F^{1...4}_c$). However, only two of them are dominant 
in the IC emission generated by HBL objects. The first dominant component ($F^1_c$) is generated 
by the low energy electrons ($K_1, N_1$) and the first part of the synchrotron spectrum ($I^1_s, 
F^1_s$). The second important component ($F^2_c$) is produced by the high energy electrons ($K_2, 
N_2$) and the first part of the synchrotron emission ($I^1_s, F^1_s$). With these 
results in mind, we can derive the formulae describing the evolution of 
emission coefficients for the dominant components
\begin{equation}
j^{1}_c(t) 
\propto K_1(t) I^1_s(t) 
\propto R_0 k_1^2 B_1 
\left(\frac{t_0}{t}\right)^{-r_e+6r_d+2r_a(n_1-1)+m(\alpha_1+1)}.
\end{equation}
\begin{eqnarray}
j^{2}_c(t) 
& \propto & K_2(t) I^1_s(t)\nonumber\\
& \propto & R_0 k_1 k_2 B_1 
\left(\frac{t_0}{t}\right)^{-r_e+6r_d+r_a(n_2-1)+r_a(n_1-1)+m(\alpha_1+1)}.
\end{eqnarray}            

By analogy to the synchrotron emission we can write the evolution of the intensity of the 
IC emission for both the discussed cases $I^{1|2}_c(t) \propto R(t) j^{1|2}_c(t).$
The evolution of the IC flux is given then by
\begin{eqnarray}
\vspace*{0.1cm}
F^{1}_c(t)  & \propto & R^2(t) I^{1}_c(t)~~\propto~~
              R_0^4 k^2_1 B_1 \left(\frac{t_0}{t}\right)^{-c_1},\\ 
              c_1 & = & 4r_e-6r_d-2r_a(n_1-1) - m(\alpha_1+1),
\end{eqnarray}
\begin{eqnarray}        
\vspace*{0.1cm}                                      
F^{2}_c(t) & \propto & R^2(t) I^{2}_s(t)~~\propto~~R_0^4 k_1 k_2 B_1 
             \left(\frac{t_0}{t}\right)^{-c_2},\\
       c_2 & = & 4r_e-6r_d-r_a(n_1-1)-r_a(n_2-1)-m(\alpha_1+1),
\end{eqnarray}
where $c_1$ describes the evolution of the IC radiation in the Thomson limit and 
$c_2$ describes the evolution of the IC emission in the Klein--Nishina regime.

\subsubsection{Components of the synchrotron self Compton spectrum: an example}

\begin{figure}[p]
\resizebox{\hsize}{!}{\includegraphics{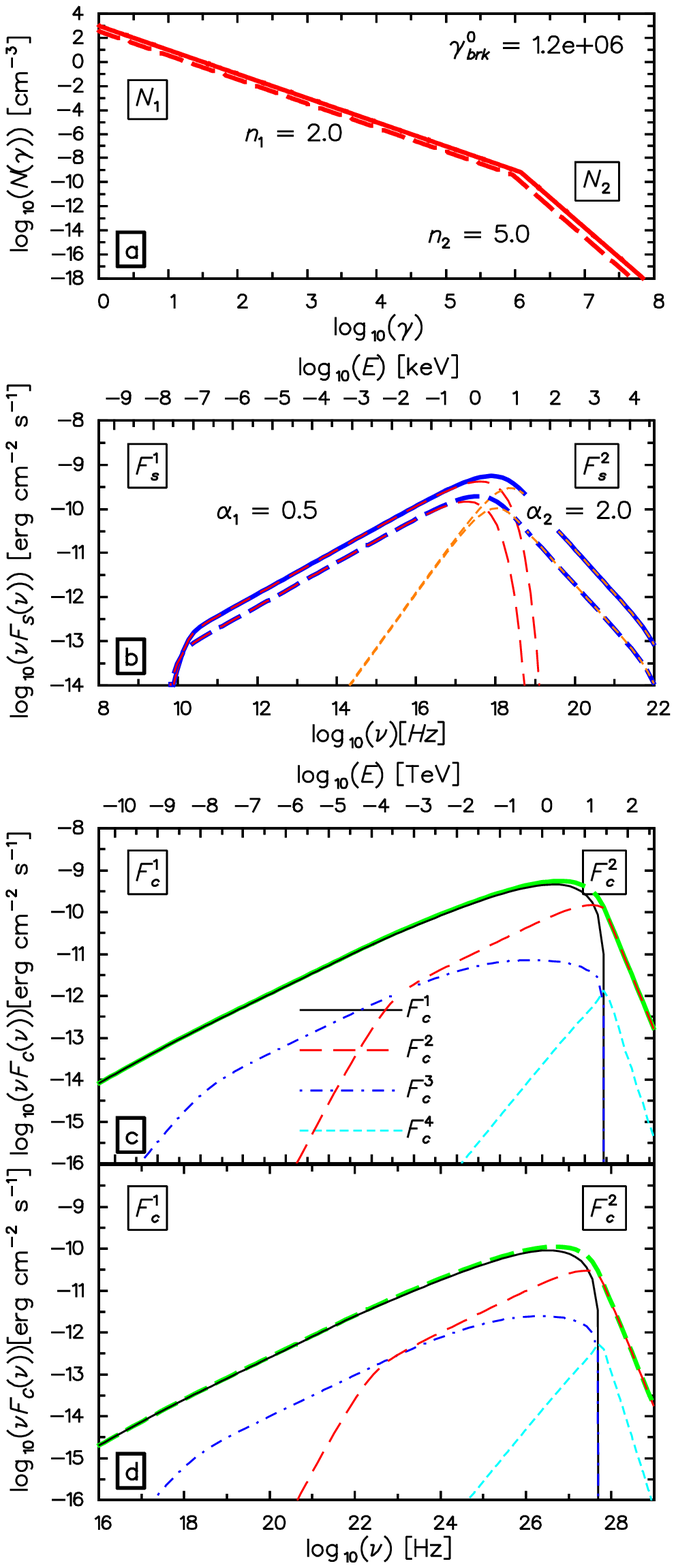}}
\caption{The evolution of the synchrotron and the IC emission of an expanding 
         spherical homogeneous source. Panel (a) shows the evolution of the 
         electron energy spectrum. The spectrum gives the synchrotron radiation 
         shown in panel (b). The evolution of the IC spectrum is shown 
         in panels (c) and (d). To simplify the figures we show only the initial 
         and the final spectra. Note that the initial spectra in all 
         figures are shown by bold--solid lines while the spectra 
         generated at the end of the simulation are shown by bold--dashed 
         lines. The basic components of the emission are shown by various thin 
         lines. A detailed description of the components and the parameters 
         used for the test are given in subsection \ref{sub_sph-geom}.
        }
\label{fig_sph-geom}        
\end{figure}

To show that the approximations used in the derivation of the equations
describing the SSC emission are correct, in Figure \ref{fig_sph-geom} we
report an example of the evolution of the synchrotron and the IC emission
generated by the expanding spherical homogeneous source, calculated using
our numerical code. The results presented in this subsection are 
correct for a wide range of values of the physical parameters usually 
assumed in order to explain the X-ray and TeV emission of the HBL objects.
However, the results are correct only for case where the dominant part of the 
synchrotron and the IC emission (the peaks) is generated by electrons 
with the Lorentz factors close to $\gamma_{\rm brk}$ which practically 
means $n_2 > 3$. The spectra presented in Figure \ref{fig_sph-geom} are divided 
into different components. This test shows that indeed only two components are
important for the high energy IC emission for a large amplitude of the
variation. Note that the components are calculated numerically,
therefore the spectra shown are slightly different from the curves
obtained from the analytical approximations by Tavecchio et
al. (\cite{tavecchio98}).  For the numerical calculations we use the SSC
emission mechanism described by Katarzy\'nski et
al. (\cite{katarzynski01}). In the test we use following parameters
$\delta = 50$, $R_0 = 2 \times 10^{16}$ [cm], $B_0 = 0.004$ [G], $k_1 =
10^3$ [cm$^{-3}]$, $\gamma_{\rm min} = 1$, $\gamma_{\rm brk} = 1.2 \times 10^6$, 
$\gamma_{\rm max} = 10^8$, $r_e = r_d = r_a = m = 1$. Note that
in an ideal case where the source is perfectly spherical and homogeneous 
and  the magnetic field inside the source is well organized the parameter 
$m$ which describes the evolution of the magnetic field should be equal to 
$2 r_e$ (e.g. $m=2$ for a linear expansion of the source) in order to 
keep the magnetic flux constant. 
However, practically the occurrence of such ideal 
sources is very unlikely. Usually we would expect a more complex geometry, 
inhomogeneity and a turbulent magnetic field inside the source which may
give a more complex evolution of the magnetic field. For example a slower 
than expected decrease of the magnetic field during the expansion may
be caused by the turbulent dynamo effect (e.g. Atoyan \& Aharonian 
\cite{atoyan99} for discussion of this problem). Therefore, we assume the 
parameter $m$ to be a free parameter in our modeling. For the linear expansion 
we assume that the value of this parameter may vary from 1 to 2. The value 
of this parameter has no influence on the results presented in this subsection. 
However, in other tests presented in the next part of this work,
we had to use $m=1$ in order to explain the quadratic correlation observed 
in the decaying phase of the flare.
Therefore, to be consistent and to provide the possibility of comparison of 
the results, in all numerical calculations presented in this paper we have 
selected $m=1$. Moreover, in this test and in all other simulations presented 
in this work we assume that the $t_0$ parameter is equal to $10~R_0/c$. This gives 
the radius expansion velocity equal to $0.1 c$ for $r_e = 1$. To change 
the source emission level at least a few times we simulate the evolution for 
a period of time equal to $3~R_0/c$.
This set of parameters is quite specific because we use a relatively small 
value of the initial magnetic field and a relatively large value of the Doppler 
factor (see estimations made by Tavecchio et al. \cite{tavecchio98} or
Katarzy\'nski et al. \cite{katarzynski01}). We assume a small value of the magnetic 
field in order to neglect radiative cooling, which cannot be described easily 
by the analytical formulae.  On the other hand, to reach the observed level of 
emission, we have to assume a large value of the Doppler factor. 
The Doppler factor value $\ge 50$ has been
used by some authors (e.g. Krawczynski et al. \cite{krawczynski02},
Konopelko et. al. \cite{konopelko03}) to explain the emission of TeV blazars.
We will prove in the next part of this work (Subsection \ref{sub_rad-cool}) that 
the radiative cooling is indeed negligible for this set of parameters.
In this test the radiative cooling is 
negligible for electrons with the energy described by the Lorentz factor 
around $\gamma_{\rm brk}$ which are producing the dominant part of the emission 
observed as the synchrotron and IC peaks. On the other hand the radiative 
cooling is still important for electrons with a Lorentz factor around 
$\gamma_{\rm max}$. However, the radiation generated by these particles is 
almost completely negligible due to the steepness of their spectrum.

Moreover, for this test and for other calculations presented in this work
we use the Hubble constant equal to 65 [km s$^{-1}$ Mpc$^{-1}$] and the
redshift equal to 0.03.  In the above calculations and in all other
calculations of the IC scattering presented in this work, we neglect the
possible absorption of the $\gamma$--ray emission due to pair production
inside the source. This process can be neglected if the Doppler factor of
the source is of the order of ten or more (e.g. Bednarek \& Protheroe
\cite{bednarek97}). Moreover, to calculate the observed TeV flux, we should 
also consider the absorption of this radiation by interaction with the Infrared 
Intergalactic Background. However, in this particular case, we are mainly 
interested in investigating the evolution of the TeV radiation with respect 
to the evolution of the synchrotron X--ray emission. Therefore, absorption 
of the TeV emission by the IR background, which is constant in time, does 
not modify the TeV/X--ray flux correlation.

\subsection{Basic cases}

Correlating the X--ray and TeV emission of the HBL objects we can 
distinguish two general cases. In the first case we compare the X--ray 
emission before the synchrotron peak ($F^1_s$) with the $\gamma$--ray 
radiation above the IC peak ($F^{2}_c$, e.g. Mrk~501, Figure~\ref{fig_501corr}). 
In the second case we correlate the X--ray emission above the synchrotron 
peak ($F^2_s$) with the $\gamma$--ray radiation above the IC peak 
($F^{2}_c$, e.g. Mrk~421, Figure~\ref{fig_421corr}). To simplify the 
discussion presented in this work we hereafter call these correlations 
$c_2/s_1$ and $c_2/s_2$ respectively. We also present in this section
the estimates of other possible correlations ($F^1_s$ vs $F^1_c$ and 
$F^2_s$ vs $F^1_c$). However, we will not discuss them in detail.
The more complex cases of the TeV/X--ray correlation, where the considered 
bands correspond to the emission around the peaks, discussed
in subsection \ref{sub_peak-corr}. 

For the tests presented in this section we use the same values of the
physical parameters as in the calculations presented in the previous 
subsection. To simulate different evolution scenarios we change only
four model parameters $r_e, r_d, r_a, m$ setting 0 or 1 as the value for 
these parameters. This means that in principle we can study only expansion of 
the source. However, as long as we neglect the radiative cooling, the 
results of such tests are opposite to the results which we could obtain 
from the compression of the source. Having four parameters and two possible 
values for these parameters (0 or 1) we have in principle sixteen different 
combinations~($2^4$). However, one combination is not interesting 
($r_e=r_d=r_a=m=0$) and four other combinations are not realistic 
(all cases where we should calculate the adiabatic cooling 
without a decrease of the particle density). In Table~\ref{tab_corr_spher} 
we show eleven other possible combinations plus one test where we 
use $m = 2$. 

We discuss here in detail only a few tests which we consider to be the most 
realistic:

\begin{itemize}

\item[$\bullet$]{
In our first case (Table~\ref{tab_corr_spher}--$a$) we assume the expansion 
of the source volume ($r_e = 1$) with constant particle density and  
magnetic field intensity ($r_d=m= 0$). This scenario could correspond to a constant 
injection of particles (e.g. from  a shock region) which extends the 
dimension of the source instead of increasing the particle density. The 
synchrotron radiation in this case is proportional to the source volume.
Therefore, the synchrotron emission produced by the low and the high energy 
electrons increases with time as $R^3 \propto t^{3r_e}$ ($s_{1|2}= 3$. 
Eqs~\ref{equ_sph_fsyn1}, ~\ref{equ_sph_fsyn2}). The IC radiation in this 
test is also proportional to the source volume. However, the IC emission is 
also proportional to the intensity of the synchrotron emission 
(Eq.~\ref{equ_sph_isyn12}) which grows linearly with the radius of the 
source ($I^{1|2}_s \propto R j^{1|2}_s \propto t^{r_e}$). Therefore, the IC 
radiation increases in time as $t^{3r_e + r_e}$ ($c_{1|2}=4$). Finally we 
obtain $c_{1|2}/s_{1|2} = 4/3$ for all possible correlations.
}

\item[$\bullet$]{
The second case ($b$) is opposite to the scenario just described. We increase 
the particle density ($r_d = 1$) keeping the volume and the magnetic 
field intensity constant ($r_e = m = 0$). A linear increase of the density provides 
a linear increase of the synchrotron radiation. The IC emission is proportional 
to the density of the particles and to the intensity of the synchrotron 
radiation which is also proportional to the density. Therefore, the IC emission 
is proportional to the square of the particle density. This is well-known 
relation. In this particular case we obtain a quadratic correlation in all 
possible cases ($c_{1|2}/s_{1|2} = 2/1$).
}

\begin{table}[!t]
\caption{Estimated correlation between evolution of the synchrotron emission and the
         evolution of the IC emission for different evolutionary scenarios.  
         Columns 2--5 show the parameters used for the tests. 
         Columns 6--9 show the estimated values of the correlation in four possible
         cases ($c_{1|2}/s_{1|2}$).
        }
\label{tab_corr_spher}        
\begin{center}
\begin{tabular}{ccccc|llll}
\hline
\hline
case  &$r_e$ & $r_d$ & $r_a$ & $m$ & $c_1/s_1$ & $c_1/s_2$ & $c_2/s_1$ & $c_2/s_2$\\
\hline
$a$& 1 & 0 & 0 & 0 & 1.333 & 1.333 & 1.333 & 1.333 \\
$b$& 0 & 1 & 0 & 0 & 2     & 2     & 2     & 2     \\
$c$& 1 & 1 & 0 & 0 & inf   & inf   & inf   & inf   \\
$d$& 1 & 1 & 1 & 0 & 4     & 1     & 7     & 1.75  \\
$e$& 1 & 1 & 1 & 1 & 2.2   & 0.786 & 3.4   & 1.214 \\
$f$& 0 & 0 & 0 & 1 & 1     & 0.5   & 1     & 0.5   \\
$g$& 0 & 1 & 1 & 0 & 2     & 1.143 & 2.75  & 1.571 \\
$h$& 0 & 1 & 1 & 1 & 1.727 & 0.950 & 2.273 & 1.250 \\
$i$& 1 & 1 & 0 & 1 & 2.332 & 1.167 & 2.333 & 1.167 \\
$j$& 1 & 0 & 0 & 1 & 1.667 & inf   & 1.667 & inf   \\
$k$& 0 & 1 & 0 & 1 & 1.667 & 1.250  & 1.667 & 1.250 \\
$l$& 1 & 1 & 1 & 2 & 1.75  & 0.7   & 2.5   & 1     \\
\hline
\end{tabular}
\end{center}
\end{table}

\item[$\bullet$]{
In the next case ($d$) we assume an increase of the volume, decrease 
of the particle density and with adiabatic cooling ($r_e=r_d=r_a=1$) with a 
constant value of the magnetic field intensity ($m=0$) during the evolution. 
Even if it is rather unlikely that the magnetic field can remain constant,
we describe this case to provide a sort of introduction to our last, more 
complex test. The synchrotron emission decreases in this test due to
the adiabatic losses because the decrease of the density is fully 
compensated for the increase in the volume if $r_e=r_d$. This decrease 
depends on the slope ($n_{1|2}$) of the particle spectrum (Eq.~\ref{equ_sph_pow_elec_spec}).
The synchrotron emission produced by the low energy electrons decreases 
more slowly ($t^{-r_a(n_1-1)}$, $s_1 = -1$) than the emission generated by the 
high energy electrons ($t^{-r_a(n_2-1)}$, $s_2 = -4$). The IC emission 
also decreases in this test. 
In the Thomson limit, the IC flux is proportional to the increase
in the volume ($t^{3r_e=3}$) and the radius ($t^{r_e=1}$, which comes from 
the intensity of the synchrotron emission $I^{1}_s \propto R j^{1}_s \propto 
t^{r_e}$) and also to the square of the density of the low energy electrons 
including the adiabatic losses ($K_1^2 \propto t^{-6r_d-2r_a(n_1-1) 
= -8}$). This finally gives $c_1 = -4$. In the Klein--Nishina regime the IC 
radiation is proportional to the evolution of the volume and radius as in the 
previous case ($t^{3r_e+r_e=4}$). However, this time the scattering is 
proportional to the density of the low energy electrons including the 
adiabatic losses ($K_1 \propto t^{-3r_d-r_a(n_1-1)=-4}$, which comes from 
the intensity of synchrotron radiation $I^1_s$) and is proportional to 
the density of the high energy electrons including the adiabatic losses 
($K_2 \propto t^{-3r_d-r_a(n_2-1)=-7}$). As a result, the IC flux in the 
Klein--Nishina regime decreases as $t^{-7}$ ($c_2 = -7$). The influence 
of the adiabatic losses provides in this test four different correlations
$c_1/s_1 = 4$, $c_1/s_2 = 1$, $c_2/s_1 = 7$, $c_2/s_2 = 1.75$.
}

\item[$\bullet$]{
In our last test ($e$) we assume an increase of the source radius, the decrease 
of the particle density with adiabatic cooling and also decay of the magnetic 
field ($r_e=r_d=r_a=m=1$). The decrease of the magnetic field causes a much 
faster decrease of the synchrotron emission ($t^{-r_a(n_{1|2}-1)-m(\alpha_{1|2}+1)}$, 
$s_1 = -2.5$, $s_2 = -7$) in comparison to the previous case. Moreover, this 
decrease increases the difference between the synchrotron emission produced by low 
and high energy electrons. The IC emission decreases faster as well. However, 
the factor ($t^{-m(\alpha_{1}+1) = -1.5}$) which marks the difference to the 
evolution discussed in our previous test is the same for the emission in the 
Thomson and in the Klein--Nishina regime. This is a consequence of the fact 
that the $\gamma$--rays in the Thomson and Klein--Nishina regime are produced 
by the scattering of the synchrotron radiation field produced by the low 
energy electrons ($I^1_s$). Therefore, the decay of the magnetic field does not 
modify the difference between the two scattering regimes. 
In conclusion, we have to add the factor --1.5 (with respect to the previous case)
to the indices describing the evolution of the IC emission, 
to obtain $c_1 = -5.5$ and $c_2 = -8.5$. 
Also in this case we obtain four different values for the correlations $c_1/s_1 = 
2.2$, $c_1/s_2 = 0.786$, $c_2/s_1 = 3.4$, $c_2/s_2 = 1.214$.
}

\end{itemize}

For all estimations presented in this subsection we have 
assumed constant values for the parameters $n_1$ and $n_2$. However, we have checked also 
that for the most realistic scenarios ($e$ \& $l$) and for all combinations of the $n_1$ value 
in the range from 1.5 to 2.5 and the $n_2$ value in the range from 3 to 8, the correlation $c_2/s_1$
is always more than quadratic while the correlation $c_2/s_2$ is always less than quadratic. 
We provide also a simple mathematical formulae for these realistic evolutions
\begin{displaymath}
\begin{array}{lcccl}
\vspace{0.4cm}           
c_2/s_1= & {\frac{\displaystyle 1+3n_1+2n_2}{\displaystyle 3n_1-1}} 
         & \mathrm{or} & 
         \frac{\displaystyle 1+2n_1+n_2}{\displaystyle 2n_1} & \textrm{TeV vs soft X--rays} \\
c_2/s_2= & \underbrace{\frac{1+3n_1+2n_2}{3n_2-1}}_{m=1}            
         & \mathrm{or} & 
         \underbrace{\frac{1+2n_1+n_2}{2n_2}}_{m=2} & \textrm{TeV vs hard X--rays} \\ 
\end{array}
\end{displaymath}

\subsection{Problem with ``the quadratic decay''}
\label{sub_quadr-prob}

This concerns the quadratic results of the $c_2/s_2$ 
correlations during the decay phase of the flare. Within the scenarios proposed in the
previous subsection there is only one solution that may explain exactly the quadratic 
correlation (see column marked $c_2/s_2$ in Table~\ref{tab_corr_spher}). However, this well-known 
test assumes only variations of the particle density. Therefore, this scenario can be 
easily used to explain the rising phase of the flare, where the activity is generated for instance 
by the injection of the particles into a constant volume of the source. We cannot easily explain the 
decrease of the density during the decay of the flare. In principle such a decrease could 
be related to the escape of the particles into a region where the magnetic field intensity
is significantly less than inside the source. However, this process does not guarantee that the
efficiency of the total IC emission will decrease two times faster than the efficiency of the 
synchrotron radiation. The IC scattering may also occur efficiently in the region where the
magnetic field is significantly less, especially close to the source surface (distance $\lesssim 
1.5R$), where the radiation field energy density is on average only two times less than on
the source surface (Gould \cite{gould79}).
The decrease of the density 
could be also related to the expansion of the source. However, assuming the expansion we have to 
consider also the influence of some additional physical process. We have to consider
the adiabatic cooling of the particles and probably also the decay of the magnetic field. All 
these processes 
can destroy the quadratic correlation which is related to the variation of the density. 
Each additional physical process which can modify the evolution 
of the synchrotron and the IC emission ($F_{s|c} \propto t^{s|c}$) in the same way ($t^{s|c+x}$) 
can destroy this correlation. 
This is a consequence of the simple fact that if $c/s = 2$ then $(c+x)/(s+x) \ne 2$. 
If the influence of the additional process is different for the evolution 
of the synchrotron emission ($t^{s+x}$) and for the evolution of the IC radiation ($t^{c+y}$) 
then we have situations where $(c+y)/(s+x)$ should give 2 if we want to keep the quadratic 
correlation. 
Therefore, assuming $s=-1$ and $c=-2$, which is a simple consequence of the decrease 
of the density, we obtain a simple relation $y=2x$ which should be fulfilled by the additional 
physical process to keep the quadratic correlation. 
For example the adiabatic cooling gives 
$x=-r_a(n_2-1)$ for the evolution of $F^2_s$ and  $y=-r_a(n_1-1)-r_a(n_2-1)$ for the evolution 
of $F^{2}_c$. Therefore, the relation $y=2x$ can be fulfilled only if $n_1=n_2$ which does not 
provide the correct solution. The efficiency of the synchrotron emission depends among other 
things on the slope of the particle spectrum and the intensity of the magnetic field. 
Therefore, 
the decay of the magnetic field gives $x=-m(\alpha_2+1)$ for the evolution of $F^2_s$ and
$y=-m(\alpha_1+1)$ for the evolution of $F^{2}_c$. Therefore, the relation can be fulfilled
if $\alpha_1 = 2\alpha_2$ which also does not give the correct solution. 
Moreover, we have
to stress that the decay of the magnetic field always results in a much faster decay of
the second part of the synchrotron radiation than the decay of the IC emission in the
Klein--Nishina regime and the Thomson limit as well. In principle this unwanted effect can be 
compensated by the adiabatic cooling which increases the decrease speed of the $F^{2}_c$ 
flux in comparison to the decrease of the $F^2_s$ flux. However, detailed 
computations show that the adiabatic cooling cannot fully compensate the influence of the decay
of the magnetic field; compare the correlations for the cases $d$ ($r_e=r_d=r_a=1, m=0$), 
$e$ ($r_e=r_d=r_a=m=1$ and $l$ ($r_e=r_d=r_a=1, m=2$) in Table \ref{tab_corr_spher}. 

\subsection{Change of the Doppler factor}

It is possible that in TeV sources the Doppler factor $\delta$ changes during 
a single flare. Evidence for this come from the requirement of a large $\delta$--value 
for the inner parts of the jet, generating most of the emission ($\delta \ge 10$, e.g. 
Dondi \& Ghisellini \cite{dondi95}, Tavecchio et al. \cite{tavecchio98},
Kino et al. \cite{kino02}, Ghisellini et al. \cite{ghisellini02}, Katarzy\'nski et al. 
\cite{katarzynski03}, or even $\delta\sim 50$, Krawczynski \cite{krawczynski02}, 
Konopelko et al. \cite{konopelko03}), while instead at the VLBI scale ($\sim$sub--pc, 
including satellite VSOP data) there is no evidence of superluminal motion of the jet's 
components (e.g. Piner et al. \cite{piner99}, Edwards \& Piner \cite{edwards02}, 
Piner \& Edwards \cite{piner04}, Giroletti et al., \cite{giroletti04}). 
This suggests a strong deceleration of the source.

To analyze the impact of the change of the Doppler factor for the observed TeV/X--ray correlation
we assume the following evolution for $\delta(t)$
\begin{equation}
\delta(t) = \delta_0 \left( \frac{t_0}{t} \right)^d,
\end{equation}
where $d$ is a free parameter. The Doppler boosting effect may change the level of the emission
[$F(\nu) \propto \delta^3 F'(\nu')$] and the observed frequency ($\nu \propto \delta \nu'$). 
For a power law spectrum ($F' \propto \nu'^{-\alpha}$) we then have
\begin{equation}
F(\nu, t) = \delta(t)^{3+\alpha} F'(\nu,t) \propto \delta_0 t^{-d(3+\alpha)} F'(\nu,t)
\end{equation}
Note that the
evolution of the flux at given frequency depends not only on the value of the Doppler factor,
but also on the spectral index $\alpha$. Therefore,
the evolution of the synchrotron emission can be different below and above the peak 
($F^{1|2}_{\rm s} \propto \nu^{-\alpha_{1|2}}$). Also the evolution of the IC radiation
depends on the spectral index ($F^{1|2}_{\rm c} \propto \nu^{-\alpha_{1|{\rm KN}}}$, where 
$\alpha_{\rm KN} \sim 2\alpha_2 - \alpha_1$, see Tavecchio et al. \cite{tavecchio98}). 

We can describe the IC and synchrotron flux evolution due to the
change of the Doppler factor by the $y$ and $x$ coefficients respectively. 
In this way the correlation which includes the change of $\delta$ becomes 
$c'/s' = (c+y)/(s+x)$. 
The change of $\delta$ will not modify the original ($c/s$) correlation
if $y/x = c/s$. 
Values $y/x > c/s$ makes the new correlation $c'/s'$ steeper ($c'/s' > c/s$),
and values of $y/x < c/s$ makes the new correlation $c'/s'$ flatter. 
This simple estimate shows that the decrease of $\delta$
can help to solve the problem of the quadratic decay,
but only slightly, since the change of the slope of the correlation
(for actual values of the synchrotron and IC spectral indices)
is only mild.

The coefficient $y$ for the evolution of the IC emission in the Klein--Nishina regime is
given by $y_2 = -3d-2d\alpha_{2}+d\alpha_1$. The parameter $x$ for the evolution of the
synchrotron emission below ($x_1$) and above ($x_2$) the peak is defined by 
$ x_{1|2} = -3d-d\alpha_{1|2}$. 
Setting $\alpha_1 = 0.5$, $\alpha_2 = 2$, and $\alpha_{\rm KN} = 3.5$,
as in the previous modeling, and assuming $d=1$, we obtain
$y_2/x_1 = 1.857$ and $y_2/x_2 = 1.3$. 
For the most realistic evolution scenarios, cases $e$ and $l$ in the Table \ref{tab_corr_spher}, 
we obtain a decrease of the $c_2/s_1$ correlation slope 
($[c_2/s_1 = 3.4 \to c'_2/s'_1 = 2.5]_e$ and 
$[c_2/s_1 = 2.5 \to c'_2/s'_1 = 2.2]_l$) and a small increase of the $c_2/s_2$ 
correlation slope 
($[c_2/s_2 = 1.214 \to c'_2/s'_2 = 1.25]_e$ and 
$ [c_2/s_2 = 1 \to c'_2/s'_2 = 1.1]_l$). 

This simple estimate shows that the impact of the change of the Doppler factor is much
stronger for the $c_2/s_1$ correlation than for the $c_2/s_2$ relation. This process does 
not modify significantly the slope of the $c_2/s_2$ correlation, therefore it cannot solve the problem 
of the quadratic decay.

\subsection{Emission around the peaks}
\label{sub_peak-corr}

\begin{figure}[!t]
\resizebox{\hsize}{!}{\includegraphics{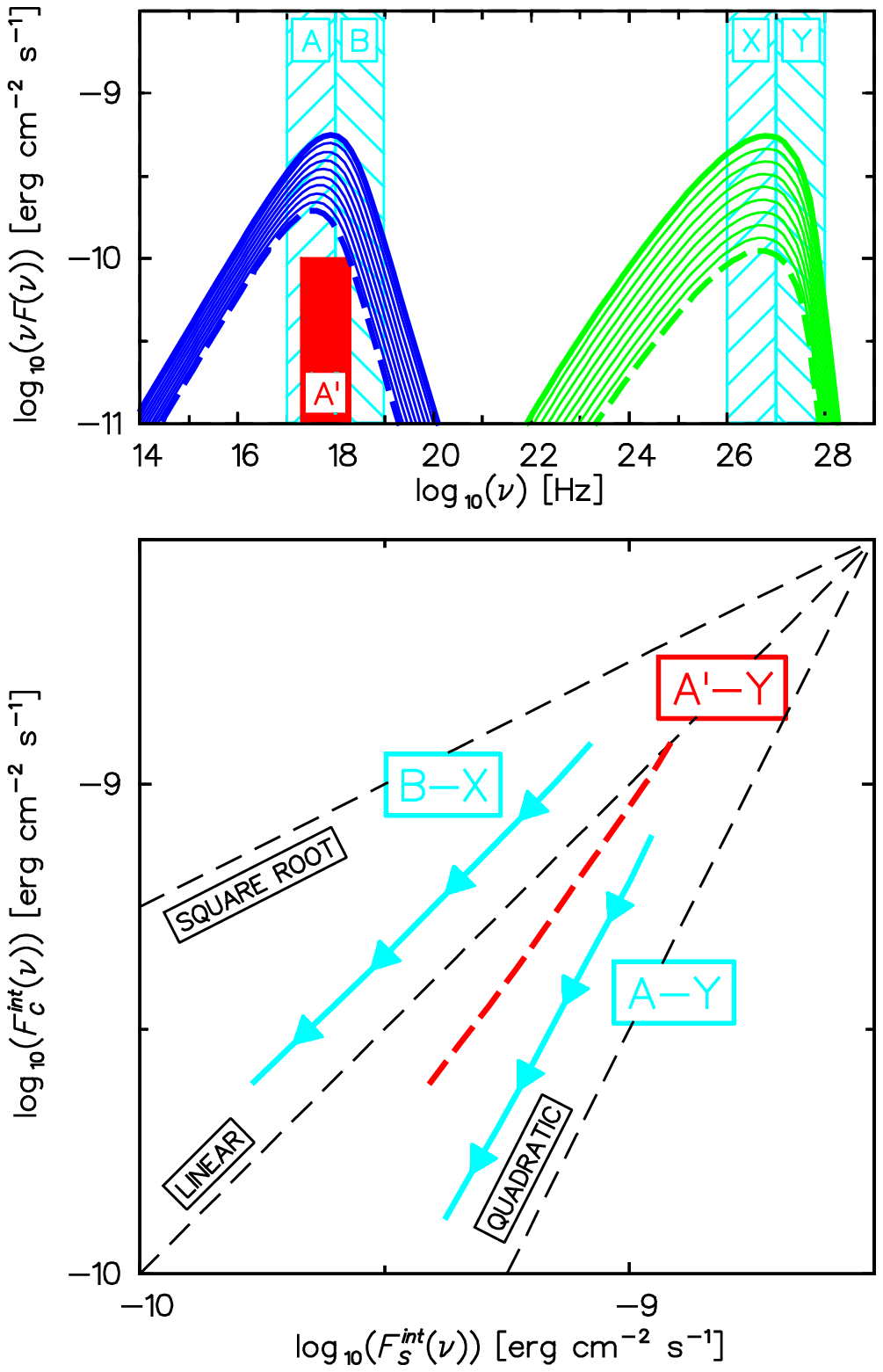}}
\caption{The upper panel shows the evolution of the SSC emission of an 
         expanding spherical homogeneous source. The areas indicated by the capital letters 
         show the spectral bands selected for the calculation of the correlation. The lower
         panel shows the two main correlations (bold solid lines with arrows which indicate 
         the direction of the evolution) calculated for the evolutions presented in the
         upper panel. The bold dashed line in the lower panel (A'-Y) shows the correlation 
         obtained after a small shift of the A--band. Thin dashed lines show a template for the 
         correlations.
        }
\label{fig_peak-corr}        
\end{figure}

Up to now, we have analyzed the evolution of the synchrotron and the IC 
emission in a case were the observed radiation belongs to the spectrum 
that can be well approximated by a power law function. In this case we 
can well describe the evolution by the simple analytical formulae 
derived in the previous subsections. Now we discuss the case where the 
emission is observed at the peaks of the $\nu F_s(\nu)$ and $\nu F_c(\nu)$ 
spectra. In this more complex case, we use the numerical 
code to calculate the correlation. Figure \ref{fig_peak-corr} shows an 
example of an evolving SSC spectrum during the expansion of homogeneous 
source. We use for this test the same set of physical parameters
used for the numerical modeling presented in subsection 
\ref{sub_sph-geom}. For each emission process we selected two spectral 
bands around the $\nu F(\nu)$ peak, named A ($10^{17}$--$10^{18}$ Hz), 
B ($10^{18}$--$10^{19}$ Hz) and X ($10^{26}$--$10^{27}$ Hz), 
Y ($10^{27}$--$10^{28}$ Hz). Note that most of the TeV emission is usually
observed in our X--band. Some observations are still possible above the 
frequency $10^{27}$ Hz (see Figures \ref{fig_501corr}-d and 
\ref{fig_421corr}-d) in our Y--band but currently we are not able to obtain 
observations in the whole Y--band. However, the spectral index in the Y--band 
is relatively steep. Therefore, radiation observed in this band is dominated 
by the emission which comes from the low frequency part of this band which 
can be observed. In our test we calculate only two correlations for 
above described bands.

The first correlation (A--Y) is calculated between the evolution of the synchrotron emission (A) 
in a transition phase (from $F^1_s$ to $F^2_s$) and the evolution of the IC radiation (Y) in 
the Klein--Nishina regime ($F^{2}_c$). This gives an almost quadratic correlation. 
This particular correlation is a kind of transition between the correlation $c_2/s_1$ (3.4 for this 
particular test) and the correlation $c_2/s_2$ (which gives 1.214 for this particular case).
The second correlation (B--X) is calculated between the evolution of the second
part of the synchrotron radiation (B, $F^2_s$) and the evolution of the IC radiation (X) in a 
transition phase from the emission in the Thomson limit to the radiation in the Klein--Nishina 
regime. The correlation in this particular case is almost linear. This is the result of the 
transition from the correlation described by $c_1/s_2$ (equal to 0.786 in this particular case) 
and the correlation $c_2/s_2$ (equal 1.214 in this particular case).

The first test presented above shows that a quadratic correlation can be obtained in 
some specific cases when we observe the radiation emitted close to the $\nu F(\nu)$ 
peaks. However, we stress that this result depends strongly on the spectral bands 
used. This effect is clearly visible in Figure \ref{fig_peak-corr} where we show the 
correlation (A'--Y) after a small shift of the A--band (A' $ \to 2 \times 10^{17}$--$ 
2 \times 10^{18}$ Hz). The almost quadratic correlation after this shift appears 
significantly less than quadratic. Moreover, to obtain the 
quadratic correlation we have assumed $m=1$. This means that the decrease of the 
magnetic field intensity was slower than we could expect from the rule which 
describes conservation of the magnetic flux ($m=2$).

\subsection{Evolution of the $n_2$ parameter}
\label{sub_n2}

In this subsection we analyse the effects related to the change of the 
index of the high energy electrons spectrum ($n_2$). The main reason for 
this test is to simulate the activity of Mrk~501 observed on April 1997.
The observations obtained at that time by the $Beppo$SAX satellite (Fig. 
\ref{fig_501corr}, Pian et. al. \cite{pian98}) indicate that the X--ray 
emission was almost stable around the energy 0.1 keV while the emission 
around 100 keV was very variable.

To simulate such evolution we change the value of the $n_2$ parameter in
ten steps, assuming $n_2=2.5$ at the beginning of the simulation and
increasing this value by a factor of 0.25 after each step. Note that 
the peak of the synchrotron spectrum, at the beginning of the test, 
is generated by the highest energy electrons ($\gamma_{\rm max} = 10^7$,
for this test) because $n_2<3$. The other model parameters are the 
same as in the previous modeling, except $\gamma_{\rm brk}$ which we 
changed to $3 \times 10^5$ in order to obtain the constant emission 
around 0.1 keV energy. The radiative cooling is negligible for 
the physical parameters used in this simulation. In the next subsection 
we discuss in detail this phenomenon and we show that indeed this 
process is negligible.

\begin{figure}[!t]
\resizebox{\hsize}{!}{\includegraphics{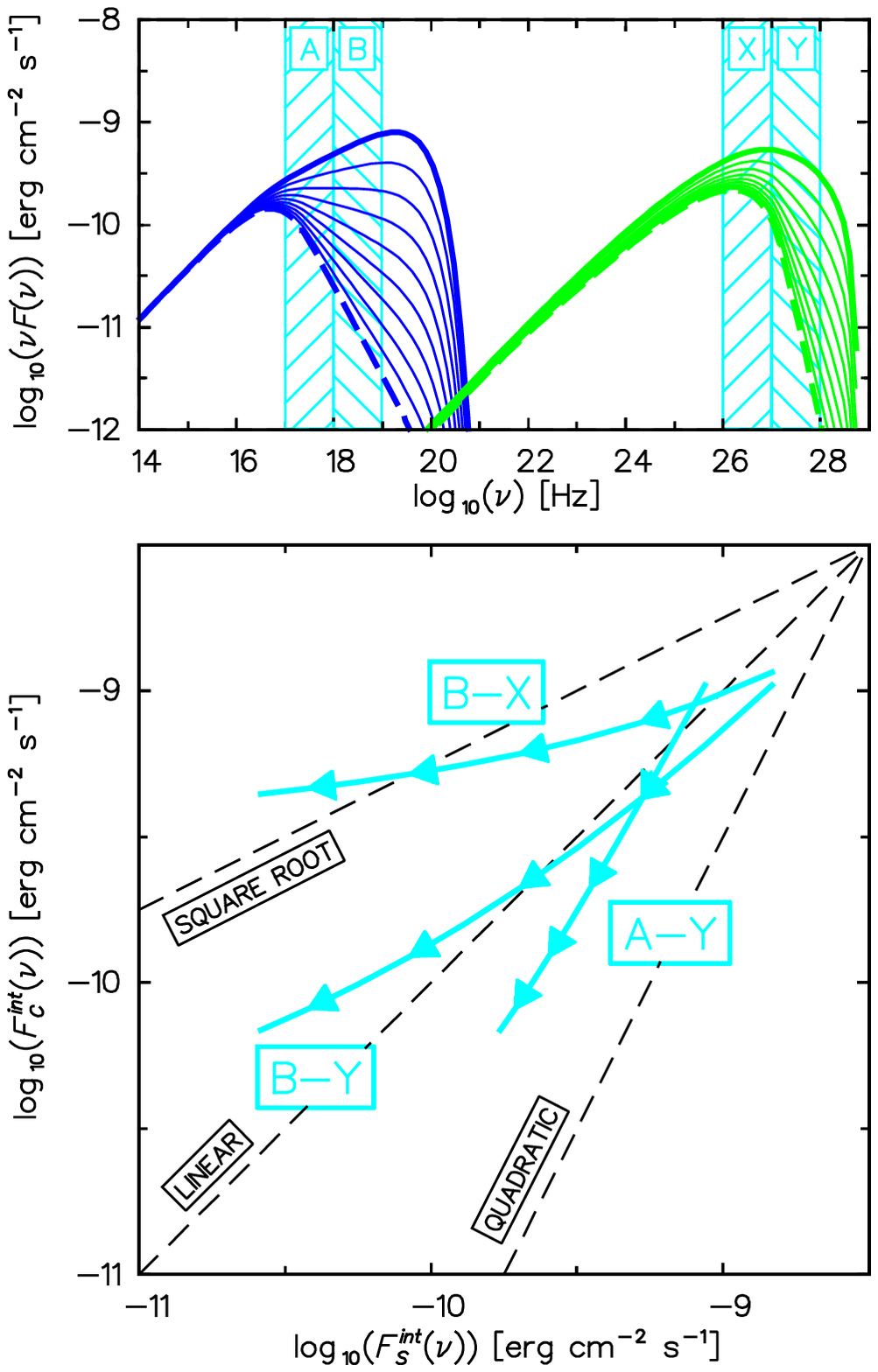}}
\caption{The upper panel shows the evolution of the SSC emission of the source where only
         the slope $n_2$ of the high energy part of the electron spectrum was modified
         during the simulation. To calculate the correlations presented in the lower panel
         we selected the same spectral bands as in the previous modeling. Thin lines in the
         lower panel show a template for the correlations.
        }
\label{fig_n2}        
\end{figure}

In the test we selected the same spectral bands for the calculation
of TeV/X--ray correlation as in the previous modeling and we also 
discus three different correlations. The first correlation (A-Y) is 
calculated selecting the TeV band decaying relatively fast and the 
part of the X--ray spectrum decreasing relatively slowly. At the 
beginning of the simulation the slope of the correlation is almost 
quadratic and becomes 1.7 towards the end of the simulation.
In the second correlation (B-Y) we selected the parts of the TeV and 
X--ray spectra decreasing relatively fast. At the beginning of the 
simulation the two fluxes vary almost linearly. However, at the end 
of the simulation the correlation slope is close to 0.5. The third 
correlation (B-X) is opposite to the first relation (a fast decay of 
the X-rays with relatively slow decay of the TeV emission) and gives 
the slope close to 0.5 at the beginning of the simulation, changing 
rapidly and reaching the value of 0.13 at the end of the simulation.

This simulation can reproduce the activity of Mrk~501 observed in April 1997. The
correlation between the Whipple and OSSE fluxes (Fig. \ref{fig_501corr}) can be explained
as the correlation between the spectral bands where the TeV and X--ray emission evolve 
relatively fast (e.g. the beginning of our correlation B-Y). The quadratic correlation between 
the Whipple and ASM fluxes can be explained as the correlation between the
part of the TeV band which evolves relatively fast and the part of the X--ray emission 
which evolves relatively slowly (e.g. our correlation A-Y). However, the 
presented examples depend strongly on the position of the peaks and on the 
selected spectral bands.

\subsection{Impact of the radiative cooling}
\label{sub_rad-cool} 

In all the scenarios discussed so far we have assumed that the
radiative cooling of the electrons is negligible. 
Now we assume that radiative cooling is important and we 
discuss its impact for the TeV/X--ray flux correlation.

To calculate the evolution of the electron energy spectrum including the
radiative cooling we use the kinetic equation
\begin{equation}
\frac{\partial N^*(\gamma,t)}{\partial t} = \frac{\partial}{\partial \gamma}
\left\{ \left[ C_{\rm cool}(t) \gamma^2 + C_{\rm adia}(t) \gamma \right] N^*(\gamma,t) \right\},
\label{equ_pde}
\end{equation}
where
\begin{displaymath}
C_{\rm cool}(t) = \frac{4}{3} \frac{\sigma_T}{m_e c} \left[U_B(t) + U_{\rm rad}(t) \right],~~~ U_B(t) = \frac{B(t)^2}{8\pi}
\end{displaymath}
describes the radiative cooling and $C_{\rm adia} = r_a/t$ describes the adiabatic cooling. 
For simplicity, we assume that the radiation field energy density ($U_{\rm rad}$)
is equal to the magnetic field energy density ($U_B$). 
The initial distribution of the electron energy spectrum is given by a 
continuous broken power law
\begin{equation}
N_0(\gamma, t=t_0) = k_1 \gamma^{-n1} \left( 1 + \frac{\gamma}{\gamma^0_{\rm brk}} \right)^{n_1-n_2}.
\label{equ_ini_elec_spec_rad_cool}
\end{equation}
The solution of this kinetic equation is given in Appendix~\ref{app_solution}. 
However, this solution must be converted to a unit volume $N(\gamma,t) = N^*(\gamma,t),
\left( t_0/t \right)^{3r_e}$ to be useful for the calculation of the emission coefficient 
(e.g.~Kardashev \cite{kardashev62}). Since the evolution of the electron spectrum is
quite complex,  we use the numerical code to check the influence of the radiative 
cooling on the correlation.

\begin{figure}[t!]
\resizebox{\hsize}{!}{\includegraphics{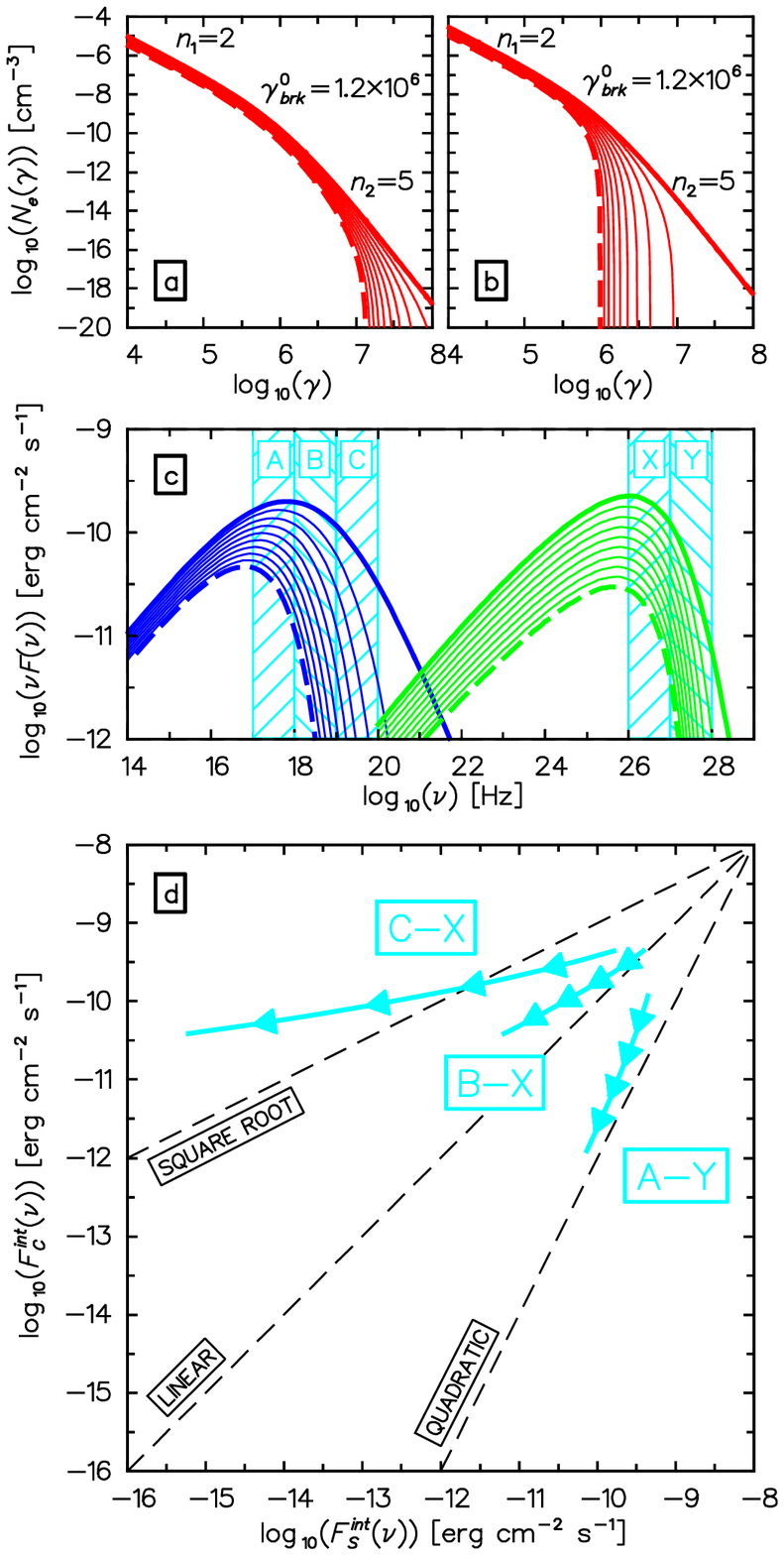}}
\caption{The impact of the radiative cooling for the correlation. Panel (a) shows the evolution 
         of the high energy part of the electron energy spectrum for the physical parameters 
         which we have selected for the modeling in the subsection \ref{sub_sph-geom}. 
         Panel (b) shows the
         evolution of the electrons spectrum for the parameters used in the subsection 
         \ref{sub_rad-cool}. The evolution of the synchrotron and the IC emission is shown in 
         panel (c). We also show the spectral bands selected for the calculation of 
         the correlations. Panel (d) shows three different correlations calculated for 
         the tested evolution. 
        }
\label{fig_rad-cool}        
\end{figure}

The radiative cooling may cause a fast decay of the high energy electrons. 
This appears as a cut--off in the high energy part of the electron spectrum. 
In Figure \ref{fig_rad-cool} we show two examples for the evolution of the spectrum. 
For the first assumed evolution (Fig.~\ref{fig_rad-cool}--a) 
the cooling is almost negligible while in the second example (Fig.~\ref{fig_rad-cool}--b) 
the cooling significantly modifies the high energy part of the electron spectrum. In the 
first test we use the same values of the physical parameters which we used for 
numerical calculations presented in the subsections \ref{sub_sph-geom}, \ref{sub_peak-corr}
and \ref{sub_n2}. In the second test we use significantly different values. 
We assume a larger initial intensity of the magnetic field ($B_0 = 0.004 \to 0.02$ [G])
to increase the radiative cooling rate. We assume a smaller value of the Doppler factor 
($\delta=50 \to 30)$ and smaller initial source volume ($R_0 = 2 \times 10^{16} \to 1.2 
\times 10^{16}$ [cm]) to keep the same level of the emission. Moreover, we also modify 
the initial density of the particles ($k_1 = 10^{3} \to 3 \times 10^3$ [cm$^{-3}$]).

Figure~\ref{fig_rad-cool}--c shows the evolution of the synchrotron and the IC spectrum 
produced by the electron spectrum given in Figure~\ref{fig_rad-cool}--b. The cut--off 
in the electron spectrum produces an exponential decay of the synchrotron and the 
IC emission. Figure \ref{fig_rad-cool}--c also shows the special bands selected for 
the calculation of the correlations. In this test we have introduced one more X--ray 
band (C, $10^{19}$ -- $10^{20}$ [Hz]) to check the correlation when using the 
exponentially decaying part of the synchrotron emission.

The first correlation (A--Y), calculated for this test (Fig. \ref{fig_rad-cool}--d), 
is analogous to the A--Y correlation computed neglecting radiative cooling. 
This is the correlation between the X--ray emission in the transition phase 
($F^1_s \to F^2_s$) and the IC radiation in the Klein--Nishina regime. 
It gives, as in the previous test, an almost quadratic result. 
The second correlation B--X was calculated between the second part of the 
synchrotron emission ($F^2_s$) and the IC radiation in the Klein--Nishina 
regime ($F^{2}_c$, $c_2/s_2$). In principle this correlation should give a 
value of the correlation around 1.2 (see Table \ref{tab_corr_spher}). However, 
the radiative cooling significantly modifies the evolution of the X--ray spectrum 
in the B--band. Thus, the X--ray emission decreases faster than in the evolution 
of the power law spectrum discussed in the previous subsection (see Figure 
\ref{fig_peak-corr}). Therefore, we obtain an almost square root correlation 
in this case. The last correlation (C--X) is the most extreme case where we 
compare the evolution of the X--ray emission in the exponential decay part (C) with the evolution
of the IC radiation in the Klein--Nishina regime ($F^{2}_c$). 
The exponential decay causes a very fast decrease of the synchrotron emission. 
Since this decrease is more than two times faster than the decrease of the IC 
radiation, the slope of the correlation is less than a square root.

The correlations presented in this subsection depend strongly on the position 
of the spectral bands. Moreover, the radiative cooling which causes the exponential 
decay of the spectra can destroy possible quadratic or linear correlations. This is 
related to the fact that the synchrotron emission responds immediately to the 
change of the high energy part of the electron spectrum while the relevant IC 
radiation is generated in the Klein--Nishina regime, mainly by the electrons with 
$\gamma$ less than $\gamma_{\rm brk}$ (see Figure \ref{fig_sph-geom}-c-d). 
Moreover, the radiation field used in the scattering is not affected by the
radiative cooling. To obtain almost quadratic result for the A-Y correlation 
we had to assume $m=1$.

\section{Cylindrical source - pizza or spaghetti}
\label{sec_cyl-src}

So far we investigated the evolution of a spherical source. The geometry of the source is
not important for the synchrotron emission as long as the electron self--absorption is neglected.
On the other hand the geometry of the source may be very important for the IC emission
which depends strongly on the radiation field available for the scattering. If the source is
strongly extended in one dimension (e.g. $x$) then only the local radiation field, proportional
to the other dimensions of the source ($y, z$), is important for the scattering. This effect
is important for the observed correlation especially if the source expands in only one or two
dimensions. To investigate this effect we approximate the emitting region by a 
cylindrical geometry, described by the radius ($R$) and the length ($L$) of the cylinder. 
We use this specific geometry to investigate two opposite cases. In the first case 
we assume $L \ll R$ a ``pizza like'' geometry (hereafter we call this the {\it pizza case}). 
In the second case we use the opposite assumption $L \gg R$ a ``spaghetti like'' geometry 
(hereafter we call this the {\it spaghetti case}). 

The main reason for such assumptions is that we are looking for robust solutions which could
explain a more than linear (or even quadratic) correlation during the decay of a flare. 
We also investigate when such solutions are independent of the specific spectral bands 
selected for the correlation. 
Using the cylindrical geometry we can simulate the expansion of the source in one dimension. 
This reduces the influence of the adiabatic cooling and the decay of the
magnetic field. Moreover, with this 
geometry we can use a lower value of the Doppler factor (e.g. $\delta \sim 25$) and 
we are still able to avoid the impact of the radiative cooling ($B_0 \sim 0.004$ [G]). 
In addition, for some specific viewing angles, we can neglect the light crossing time effect
(e.g. a pizza--like source viewed face on, or a spaghetti--like source viewed at
$1/\Gamma$, corresponding to a viewing angle of $90^\circ$ in the comoving frame).

We assume that the radius and the length of the cylindrical source evolve in time as  
power law functions
\begin{equation}
R(t) = R_0 \left(\frac{t_0}{t} \right)^{-r_e},~~~~~
L(t) = L_0 \left(\frac{t_0}{t} \right)^{-l_e},
\end{equation}
where $R_0$ and $L_0$ are initial radius and length respectively. This parameterization
gives a volume which evolves in time as $V(t) = \pi R_0^2 L_0 \left(t_0/t\right)^{-E}$, 
where $E = (2r_e+l_e)$. 

The initial distribution of the electron energy spectrum is assumed to be a broken power 
law function as in the previous model (see Eq.~\ref{equ_ini_elec_spec}). The
evolution of the spectrum is given by a minimum of two power law functions as in the 
previous modeling (see Eq.~\ref{equ_min_elec_spec}). However, for this modeling the 
definition of the power law functions substituted for the minimum is different 
according to the different parameterization of the cylindrical geometry. The functions 
are given by the formulae
\begin{displaymath}
\begin{array}{ll}
\vspace*{0.1cm}
N_{1|2}(\gamma, t) = K_{1|2}(t) \gamma^{-n_{1|2}}; &~~~~~K_{1|2}(t) = k_{1|2} \left(\frac{t_0}{t} \right)^{D}
                                                                \left(\frac{t_0}{t} \right)^{A(n_{1|2}-1)},\\
\end{array}                                               
\end{displaymath}
where $D = (2r_d + l_d)$ describes the evolution of the density and $A = (2r_a + l_a)/3$ 
describes the adiabatic heating or cooling of the particles.

The evolution of the magnetic field intensity for this model is defined in the same way as in 
the previous modeling (see Eq.~\ref{equ_evol_magn}).

The evolution of the synchrotron emission coefficient is calculated for the low ($j^1_s$) and 
high energy electrons ($j^2_s$) separately
\begin{equation}
j^{1|2}_s(t) \propto K_{1|2}(t) B(t)^{\alpha_{1|2}+1}
             \propto k_{1|2} B_{1|2} \left(\frac{t_0}{t}\right)^{D+A(n_{1|2}-1)+m(\alpha_{1|2}+1)}.
\end{equation}
The evolution of the intensity of the synchrotron emission is calculated along the source length
\begin{equation}
I^{1|2}_s(t) \propto L(t) j^{1|2}_s(t) \propto  L_0 k_{1|2} B_{1|2} \left(\frac{t_0}{t}\right)^{-l_e + D+A(n_{1|2}-1)+m(\alpha_{1|2}+1)},
\end{equation}
If we assume that the observer is located outside the source at some distance on the symmetry 
axis of the cylinder then the observed flux is given by
\begin{eqnarray}
\vspace*{0.1cm}
F^{1|2}_s(t) & \propto & R^2(t) I^{1|2}_s(t) ~\propto~ R_0^2 L_0 k_{1|2} B_{1|2} \left(\frac{t_0}{t}\right)^{-s_{1|2}},\\
    s_{1|2}  & =       & E - D - A(n_{1|2}-1) - m(\alpha_{1|2}+1).
\end{eqnarray}

The main difference between the previous modeling and the current calculations is 
in the IC emission. 
The difference comes from the energy density of the synchrotron
emission ($U_{rad}$) which is available for the scattering. 

First we describe the radiation in the pizza case where the evolution of the emission 
coefficient in the Thomson limit is defined by $j^{1}_{c_p}(t) \propto K_1(t)~I^1_s(t)$.
The evolution of the IC emission coefficient for the Klein--Nishina regime is given by
$j^{2}_{c_p}(t) \propto K_2(t)~I^1_s(t)$.
In this particular pizza case we use for the calculation of the scattering only local 
radiation field that is proportional to the length of the source ($U_{rad} \propto I^1_s 
\propto j^1_s L$). We assume that the contribution of the synchrotron emission from the 
other parts of the source is negligibly small because the radiation field energy density 
decreases with the square of the distance. According to this assumption the radiation
field used for the calculation of the IC scattering is isotropic. To obtain the evolution 
of the observed flux we have to multiply the evolution of the emission coefficient defined 
above by the source volume, which gives
\begin{eqnarray}
\vspace*{0.1cm}
F^{1}_{c_p}(t)  & \propto & R_0^2 L_0^2 k^2_1 B_1 \left(\frac{t_0}{t}\right)^{-c_{1P}},\\ 
         c_{1P} & =       & 2r_e + 2l_e - 2D - 2A(n_1-1)-m(\alpha_1+1),
\end{eqnarray}
for the Thomson limit and
\begin{eqnarray}        
\vspace*{0.1cm}                                      
F^{2}_{c_p}(t)  & \propto & R_0^2 L_0^2 k_1 k_2 B_1 \left(\frac{t_0}{t}\right)^{-c_{2P}},\\
\vspace*{0.1cm}
         c_{2P} & =       & 2r_e + 2l_e - 2D -A(n_1-1)-A(n_2-1)\nonumber\\
                & -       & m(\alpha_1+1),
\end{eqnarray}
for the Klein--Nishina regime.

In the spaghetti scenario we use only the local radiation field for the calculation of 
the scattering. However, now the local radiation field is proportional to the source 
radius ($U_{rad} \propto j^1_s R$). 
Therefore, the evolution of the emission coefficient in the Thomson limit is given by
$j^{1}_{c_s}(t) \propto K_1(t)~R(t)~j^1_s(t)$.
The evolution of this coefficient for the Klein--Nishina regime is defined by
$j^{2}_{c_s}(t) \propto K_2(t)~R(t)~j^1_s(t)$. Finally, the evolution of the 
observed fluxes are calculated in the same way as in the pizza scenario. For 
the Thomson limit we have
\begin{eqnarray}
\vspace*{0.1cm}
F^{1}_{c_s}(t) & \propto & R_0^3 L_0 k^2_1 B_1 \left(\frac{t_0}{t}\right)^{-c_{1S}},\\ 
        c_{1S} & =       & 3r_e + l_e - 2D - 2A(n_1-1)-m(\alpha_1+1),
\end{eqnarray}        
and for the Klein--Nishina regime we have
\begin{eqnarray}
\vspace*{0.1cm}                                      
F^{2}_{c_s}(t) & \propto & R_0^3 L_0 k_1 k_2 B_1 \left(\frac{t_0}{t}\right)^{-c_{2S}},\\
\vspace*{0.1cm}
         c_{2S} & =       & 3r_e + l_e - 2D -A(n_1-1)-A(n_2-1)\nonumber\\
                & -       & m(\alpha_1+1).
\end{eqnarray}

\subsection{Specific solutions}

With respect to the spherical geometry, the cylinder is parametrized by two 
more free parameters ($L_0$, $l_e$) which give more possible scenarios 
for the time evolution of the flux. In this subsection we discuss 
only a few possible cases that give directly a quadratic correlation between 
the second part of the synchrotron emission ($F^2_s, s_2$) and the IC 
radiation in the Klein--Nishina regime ($F^{2}_c, c_2$) during the decay of
a flare.

\begin{itemize}

\item[$\bullet$]{
Consider a pizza--like geometry and only linear increase of the source length 
($l_e=1$, $r_e=r_d=r_a=l_d=l_a=m=0$). The increase causes a linear growth 
of the synchrotron emission ($F^{1|2}_s \propto t^{l_e}, s_{1|2} = 1$) and 
a quadratic ($c_{(1|2)P} = 2$) increase of IC radiation which is proportional 
to the increase of the volume and the radiation field as well ($U_{rad} 
\propto j_s L \propto t^{l_e}$). Thus in all possible cases we can 
obtain the quadratic correlation ($c_{(1|2)P}/s_{1|2} = 2$). The same test 
for the spaghetti case gives $c_{(1|2)P}/s_{1|2} = 1$ because the local 
radiation field is proportional to the radius and is constant. Note that the 
similar test for the spherical geometry ($r_e=1, r_d=r_a=m=0$) gives 
$c_{1|2}/s_{1|2} = 1.333$.
}

\item[$\bullet$]{
In the next test we assume a pizza--like geometry with expansion of 
the radius, decrease of the density and the adiabatic losses 
that correspond to the change of the radius ($r_e=r_d=r_a=1$).
The other parameters are assumed to be constant ($l_e=l_d=l_a=m=0$). 
For $r_e=r_d$ the expansion of the volume is compensated entirely by 
the decrease of the density. Therefore, the synchrotron emission decreases
only due to the adiabatic cooling ($F^2_s \propto t^{-2r_a(n_2-1)/3}, 
s_2 = 2.67$). The IC emission in the Klein--Nishina regime depends on the 
expanding volume ($t^{2r_e=2}$), the square of the decreasing density 
($t^{-4r_d=-4}$) and the adiabatic losses of the low and high energy 
electrons ($t^{-2r_a(n_{1|2}-1)/3 = -(0.67|2.67)}$). The combination of 
the above processes gives a decrease $F^{2}_c \propto t^{-5.33}$, 
which finally provides a quadratic correlation ($c_{2P}/s_2 = 2$). 
However, the correlation depends on the value of the $n_1$ and $n_2$
parameter ($n_{1|2} = 2|5$, in the above test). We have checked that 
for a given value of the parameter $n_1$ in the range from 1.5 to 2.5 
there is always only one value of the index $n_2$ which provides the 
quadratic result (e.g. $n_1 = 2.5$ and $n_2=5.5$). Other correlations 
for this case, which we do not discuss in detail, are described by
$c_{1P}/s_1=5$, $c_{1P}/s_2=1.25$, $c_{2P}/s_1=8$.
}

\item[$\bullet$]{
The quadratic correlation can be obtained also in a three dimensional 
expansion of a pizza--like source ($r_e=r_d=r_a=l_e=l_d=l_a=1$). However, 
to get the quadratic result it is necessary to assume a constant 
magnetic field ($m=0$) during the expansion and specific values of the 
parameters $n_1$ and $n_2$ (e.g. $n_1=2$ and $n_2=4$). Note that similar 
three dimensional expansion of the spherical source ($r_e=1=r_d=r_a=1, 
m=0$) gives $c_2/s_2 = 1.75$.
}

\item[$\bullet$]{
The spaghetti scenario also provides a quadratic correlation if we assume the expansion of
the radius and the length of the source ($r_e=r_d=r_a=l_e=l_d=l_a=1$). However, also in this 
case we have to assume a constant magnetic field ($m=0$) and specific slopes of the electron
spectrum (e.g. $n_1=2$ and $n_2=4$) to get the right correlation.
}

\end{itemize}

We conclude that the cylindrical geometry can provide
some quadratic solutions if we can assume a constant or almost constant value of the magnetic
field during the evolution of the source. Moreover, we have also to assume that the impact
of the radiative cooling is negligible. Otherwise with this particular
geometry the explanation of the quadratic result of the correlation $c_2/s_2$ is rather 
problematic.

\section{Light crossing time effects (LCTE)}
\label{sec_lcte}

We distinguish two different types of LCTE. The first effect, 
which we call the internal LCTE, is important for the 
calculation of the evolution of the radiation field inside the source. This effect is 
especially important for precise calculation of the IC emission. The second effect, 
which we call the external LCTE, is important for the observer. 
We investigate in detail in this subsection this second effect, which seems to 
have a stronger impact for the correlation between the TeV and X--ray emission.
Note that recently Sokolov, Marscher \& McHardy (\cite{sokolov04})
proposed a new model to explain the rapid multifrequency variability 
of blazars. They investigated in detail the influence of LCTE for the observed
emission. This model provide especially precise description of the IC scattering 
in a case where the internal LCTE is important.

\begin{figure}[!t]
\resizebox{\hsize}{!}{\includegraphics{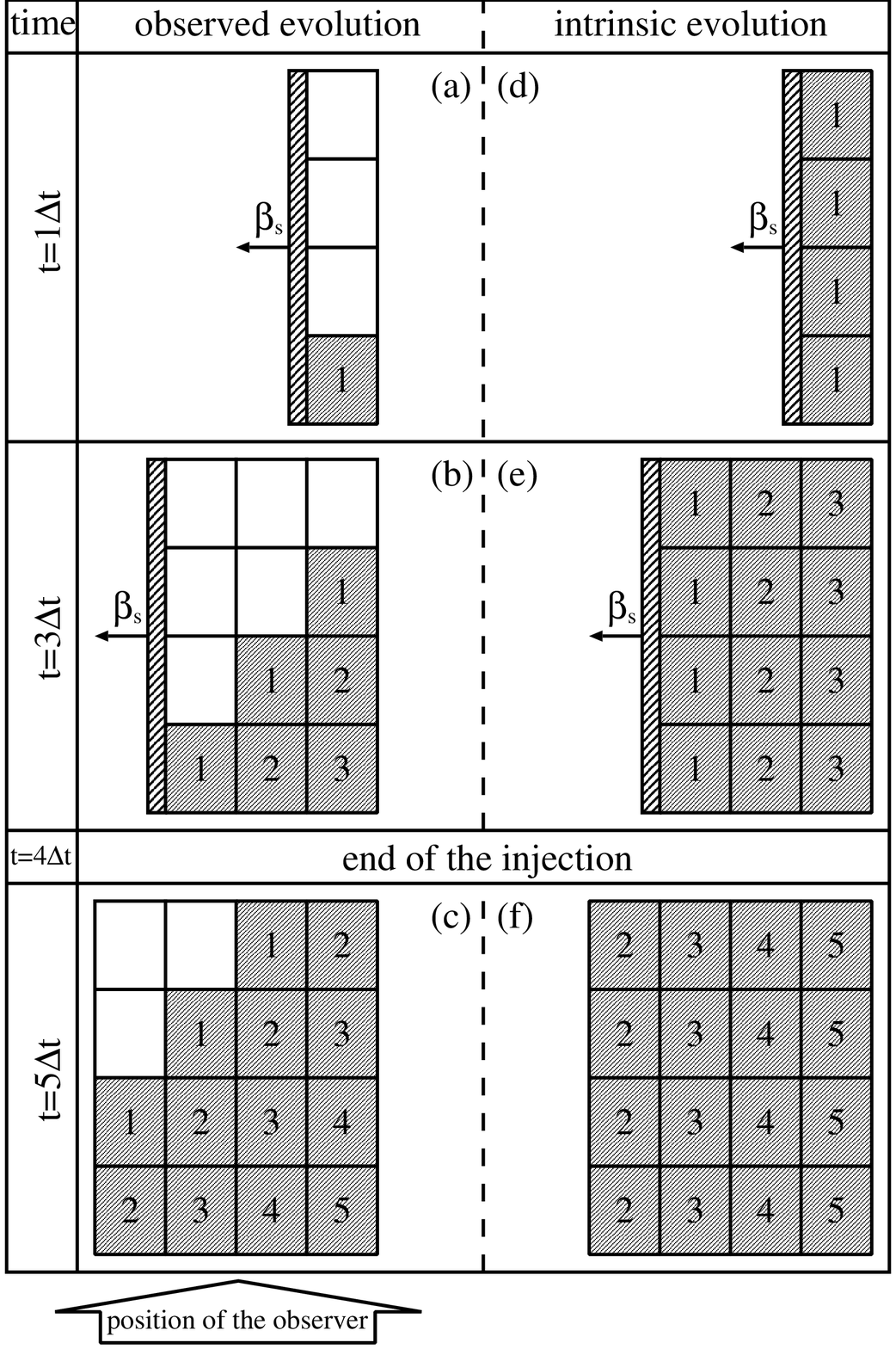}}
\caption{The evolution of the source according to the model proposed by Chiaberge $\&$ 
         Ghisellini (\cite{chiaberge99}). The right column shows evolution of the 
         total volume of the source. The left column shows the cells that
         contribute to the source emission observed at given time by the observer located at 
         $90^{\circ}$ (bottom of the figure) with respect to the shock velocity ($\beta_s$) in the
         comoving frame. The number in each cell indicates the age of the particles. Note that 
         at time $4\Delta t$ the injection process was stopped.
        }
\label{fig_lcte}        
\end{figure}

To check the influence of the external LCTE for the correlation we have used the model proposed 
by Chiaberge $\&$ Ghisellini (\cite{chiaberge99}). 
They assumed that the source is created by a shock wave which accelerates the electrons. 
The electrons are injected from the shock region into a volume of the jet where 
they generate most of the synchrotron and the IC radiation. 
The geometry of the emitting region is approximated to be cubic and the region has been 
divided into small homogeneous cells. 
In figure \ref{fig_lcte} we sketch
the evolution of such a source. 
On the right we show the intrinsic evolution. 
On the left we show the evolution of the part of the source that can 
be observed in the comoving frame at $90^{\circ}$ with respect to the shock velocity.
The number in each cell indicates the age of the particles inside the cell.

If the external LCTE is not important then the observer can 
see immediately the changes of the total source structure. 
This condition is true for this model if the shock velocity ($\beta_s$) 
is significantly less than the speed of light. 
Otherwise only part of the source volume can be observed by the observer at given time. 

We examine two theoretical scenarios of the source evolution. 
In the first we assume that the emission of a single cell is constant in time. 
For the second scenario, opposite to the previous one, 
we assume a very fast flux decay within a single cell. 

\subsection{Constant emission of a single cell}

The assumption of constant emission requires the same age of the particles 
in all cells. 
This means that in cell presented in Figure \ref{fig_lcte} we should 
have the same number--1. 
To check the influence of the process for the evolution of the emission 
we compare the intrinsic evolution of the source with the observed evolution.

For the intrinsic evolution of the source the constant injection of the cells 
gives a linear increase of the synchrotron emission, which is proportional 
to the increase of the source volume. 
If we neglect the internal LCTE, the IC radiation also will increase proportionally to the 
increase of the volume. However, this assumption is not precise and we will discuss the problem 
of the internal LCTE at the end of this section. Immediately after the the injection stops
(for example due to destruction of the shock wave) we should observe a constant emission of the 
source where the absolute level of the emission is proportional to the created volume.

If we consider the external LCTE then observer at the beginning of the evolution ($t=\Delta t $) 
will receive only the emission from the cell which is nearest to him (bottom row of the cells). 
At a later time the observer will start to receive successively emission produced by the cells 
located at larger distances. If the emission of a single cell is constant in time 
(number 1 in all observed cells) then the observed emission increases as a sum
\begin{equation}
\sum_{k=1}^n k = \frac{n(n+1)}{2},
\end{equation}
where $n$ is the number of cells in the bottom row. 
Note that this formula is correct only for the source which at the end of the 
injection is a square ($n \times n$ cells) and is correct only for the injection phase. 
This simple estimation indicates that the observed flux will increase in time 
as $t^{\sim 2}$ if $n \gg 10$. 
This result is significantly different from the intrinsic evolution where 
the increase of the emission in the injection phase was linear. 
Therefore, this simple test shows that the external LCTE may 
significantly change the evolution of the observed emission. 
Note that in this test the level of the emission increases also after the 
end of the injection. 
However, in this phase the increase is slower 
than $t^{\sim 2}$ and after some time ($2 \times l/c$ where $l$ is the 
dimension of the square source) the flux level becomes constant,
and equal to the level of the intrinsic emission after the end of the injection.

\subsection{Very fast decay of a single cell}

We now assume that the flux, within a single cell, decays very fast. 
The ``life time" of a single cell is $\Delta t$.

If the external LCTE is not important we observe constant emission of the source as long
as the shock is injecting particles. 
The total volume of the source during the whole 
injection process is represented only by the cells next to the shock front (e.g.
Figure \ref{fig_lcte}-d). The source disappears abruptly after the end of the injection.

When the external LCTE is important, at the beginning of the injection the 
observer will see only the radiation produced by the bottom row, reduced in 
this particular case to a single cell. At a later time the radiation produced 
by the cells at larger distances will arrive at the observer. 
This means that the observer will see a linear increase of the emission. 
The duration of this increase depends on the number 
of cells ($k$) in the column next to the shock front. 
After the time $k \times \Delta t$ the 
observed emission should be constant. 

After the end of the injection we should observe a
linear decrease of the emission with a duration equal to the light crossing
time from the back to the near part of the source.
This is the opposite behavior to the linear increase observed at the beginning 
of the simulation.

There are two important conclusions from this test. 
The first is that even if
the intrinsic emission of the source is constant for some time at the beginning of the
evolution, the observer will see a linear increase of the radiation. 
The second conclusion concerns the decay of the source. 
Even if the source ``dies" abruptly in a very short
time, the observer will see a linear decay of the emission for a time corresponding to
the dimension of the source. 
The conclusions obtained in this and in the previous subsection 
are strictly valid for the specific adopted geometry.
In the next subsection we discuss more possible implications of LCTE.

\subsection{LCTE - general conclusions}

The model used to explore the influence of LCTE for the evolution of the source emission 
gives the possibility to investigate only the external LCTE. However, the conclusion 
obtained from the simple test which we have performed with this model seems to be quite 
strong. If the emission produced by the total volume of the source evolves in time as $t^x$
then the observed emission evolves as $t^{x+1}$. 
Besides being confirmed by our simple analytical tests,
this very simple rule is also confirmed by numerical calculations assuming several
different decay rates of a single cell emission. 
However, this result is strictly valid only for the proposed geometry, source evolution and 
specific position of the observer (at $90^{\circ}$). 
The model used
cannot precisely describe the evolution of the radiation field inside the source 
(internal LCTE) and cannot describe the emission of the expanding source. 
This would require a more sophisticated numerical modeling.

If we assume that the external LCTE indeed increases by one the index which describes the
evolution of the emission, then we can analyze the impact of this process for the observed
correlation. 
If for example the correlation between the synchrotron and the IC emission is
intrinsically quadratic then the observed correlation will not be quadratic. 
This is a simple 
consequence of the rule already discussed in subsection \ref{sub_quadr-prob} 
(if $c/s = 2$ then $(c+x)/(s+x) \neq 2$). 
On the other hand if we observe a quadratic 
correlation it means that the intrinsic change of the IC emission must be much larger 
that the change of the synchrotron radiation. 
For example we need $F_s \propto t^{s= 1}$ 
and $F_c \propto t^{s= 3}$ intrinsically to obtain a quadratic correlation after the
basic transformation due to the external LCTE ($(c= 3+1)/(s= 1+1) = 2$). The transformation
indicates that only some specific intrinsic evolutions of the synchrotron and the IC emission
(e.g. $c=1~\&~s=0$, $c=3~\&~s=1$, $c=5~\&~s=2$) may provide a quadratic 
correlation. 
Similar arguments apply to any observed correlation slope.

The significantly faster evolution of the intrinsic IC flux with respect to the
evolution of the intrinsic synchrotron flux seems difficult to explain. 
If we neglect
the internal LCTE, then for the calculation of the IC scattering inside a single cell we
use only the synchrotron radiation field produced within this cell. 
With this assumption the evolution of the IC and synchrotron radiation will be exactly 
the same ($c=s$). 
However this approach is of course not always exact, since electrons 
inside a given cell may scatter the seed photons produced also by the surrounding cells. 
The cell at the center of the source is the one most affected by this effect. 
For such a cell we could have a linear increase of the 
radiation field during the injection phase. 
The cells affected the least will be the ones at the corners of the source 
(for them the effect should be four times less than for the cell in the center). 
Therefore, the evolution of the
intrinsic IC emission of the whole source should be more than linear but less than 
quadratic ($F_c \propto t^{1~<~c~<~2}$) which after the transformation gives $2<c<3$.
This estimation requires constant intrinsic synchrotron emission to explain a
quadratic correlation in the observer frame.

To conclude:
i) if the intrinsic evolution of the source produces a given
correlation between the synchrotron and IC emission, then the external LCTE yields
a correlation with a different slope;
ii) if we observe a given correlation and the external LCTE is important,
then the intrinsic relative IC/synchrotron change must be much stronger than 
what is observed.

\section{Summary and conclusions}

We have presented a detailed study of the expected correlation between
the variations observed in the X-ray and TeV bands in HBL, in the
context of the widely-used homogeneous SSC model. This work has been
stimulated by the observations of both linear and quadratic
correlations in the few cases for which the available data are
suitable for a detailed study of the correlation in single flaring
events (Section \ref{sec_obs-corr}).

First we addressed the problem in the context of the widely used
spherical geometry (Section \ref{sec_spher-src}), presenting analytical 
relations valid when the radiative cooling can be neglected and numerical 
results valid in general. While a quadratic correlation during the 
increasing phase of a flare is easy to reproduce through an increase
in the density (due, for instance, to a continuous injection of new
particles in the emitting volume), we found that the same correlation
observed during the decreasing phase poses a difficult problem. A case
close to a quadratic relation can be obtained only in the rather
physically implausible case of adiabatic expansion and a constant
magnetic field.  When the observational bands comprise the peaks of
the SEDs the situation is more complex even if a close-to-quadratic
relation can be obtained, the solution strongly depends on the exact
position of the bands and a small change in the observational limits
inevitably would change the correlation.  

The main conclusion that we can derive from the first part of our study
is that in order to to get a quadratic (or, even more general,
more-than-linear) correlation between X-rays and TeV we need rather
special conditions and/or fine-tuning in the temporal evolution of the
physical parameters. For special choices of the spectral bands under
study and of the parameters describing the evolution we can get even
more-than-quadratic relations. In all the cases these solutions are 
quite ``delicate'' a small change in the observational band and/or 
in the parameters inevitably changes the correlation.

Next (Section \ref{sec_cyl-src}) we investigated the changes related to
a source geometry, assuming that the emitting region is a cylinder. 
Synchrotron emission is not affected by the actual shape of the emission 
region (as long as absorption effects within the source are negligible). 
On the other hand the value of the radiation energy density strongly
depends on the geometry, affecting the SSC emission. We distinguish
between two cases (called pizza or spaghetti) depending on the ratio
between the length and radius of the cylinder ($R/L>1$ or $R/L<1$
respectively). We found that, similar to the spherical case, quadratic
solution can be found only assuming that the magnetic field is not
affected by the expansion. Moreover, radiative cooling can destroy the
correlation.

Although light crossing time effects have not been directly taken into
account in the calculations, we have briefly discussed possible effects
of this phenomena (Section \ref{sec_lcte}). We show that the external
LCTE can ``weaken'' the intrinsic (within the source) correlation. 
On the other hand if we observe for example a quadratic correlation and
the external LCTE is important then the intrinsic IC/synchrotron change 
must be stronger than quadratic (e.g. $c/s = 3/1$).

As a last step we discussed the possibility that the SSC occurs in the
Thomson regime. This would easily produce a quadratic correlation
between X-rays and TeV, since the seed photons for the IC scattering
are produced by the same electrons. However this would require rather
implausible conditions for the emission region (extreme Doppler
factors, $\delta\sim 1000$, and an extremely compact source, see Appendix
\ref{app_thomson}).

The general conclusion is that if the quadratic correlation during
single flares will be found to be common, the simple homogeneous SSC
model will face a severe problem. As stressed many times,
the main difficulty is to explain the decaying phase, therefore
acceleration processes and/or particular injection mechanisms cannot
contribute to solving the problem.

\begin{acknowledgements}
We are grateful to E. Pian, A. Djannati-Atai, M. Catanese and H.
Krawczynski for the data obtained by the $Beppo$SAX, CAT, Whipple, OSSE, 
RXTE-PCA and HEGRA experiments. This article made use of observations
gathered by RXTE/ASM experiment, provided by HEASARC, a service of 
NASA/Goddard Space Fight Center. We acknowledge the EC funding under 
contract HPRCN-CT-2002-00321 (ENIGMA network).
\end{acknowledgements}

\appendix

\section{Large Doppler factor -- the Thomson limit}
\label{app_thomson}

The main difficulty in explaining a more than linear correlation between 
the TeV and the X--ray fluxes is associated with the fact that in TeV 
sources the seed photons to be scattered at high energies are those below 
the peak of the synchrotron emission, because the Klein--Nishina decline 
of the scattering cross section with energy implies a small efficiency for
the scattering between photons at the synchrotron peak and electrons
with random Lorentz factor close to $\gamma_{\rm brk}$. This
``decoupling'' between scattering electrons and scattered seed photons
introduces the problems we have discussed.

Indeed, Tavecchio et al. (\cite{tavecchio98}) have demonstrated that 
in the single zone, homogeneous SSC model, the knowledge of the peak 
frequencies $\nu_s$ and $\nu_c$, the flux at these frequencies, and the
variability timescale is sufficient to determine the physical parameters 
of the source.  In this case one associates the
variability timescales $t_{\rm var}$ with the size $R$ of the source,
through $R=ct_{\rm var}\delta$.  Application to Mrk~421 and Mrk~501
using this prescription results in the finding that both sources are
suffering from Klein--Nishina effects.

We now relax the assumption $R=ct_{\rm var}\delta$, and require
instead that the scattering of X--ray photons with the electrons
producing these same photons (by the synchrotron process) occurs in
the Thomson limit.  This will have the advantage that any variations
of the number of electrons emitting in the X--ray band will
automatically produce a quadratic variation in the TeV band.

To this aim we define $\gamma_x$ as the random Lorentz factor of those
electrons producing synchrotron photons of frequency $\nu_x$, and
require (Thomson limit)
\begin{equation}
{h\nu_x\over m_ec^2}\, { \gamma_x\over \delta} \, < \, 1
\label{kn}
\end{equation}
Since $\nu_x \simeq 3.6\times 10^6 B\gamma_x^2 \delta$, the above
condition yields
\begin{equation}
B\delta^3 \, > \, \left({h \over m_ec^2}\right)^2 \, 
{ \nu_x^3 \over 3.6 \times 10^6}
\label{bd3}
\end{equation}
If the scattering occurs mainly in the Thomson limit, then 
the ratio between the self Compton and the synchrotron
peak frequencies is $\nu_c/\nu_s=(4/3)\gamma_{\rm brk}^2$.
Since the synchrotron peak frequency is given by
$\nu_s \simeq  3.6\times 10^6 B\gamma_{\rm brk}^2 \delta$,
we have
\begin{equation}
B\delta \, = \, {\nu_s^2 \over 2.8 \times 10^6 \nu_c}
\label{bd}
\end{equation}
The third condition concerns the synchrotron and self Compton 
peak fluxes. 
The intrinsic (comoving) synchrotron radiation energy density 
is given by $U^\prime_{\rm syn} \, = \, {L_{\rm syn} / 4\pi R^2 c \delta^4}$.
Since, in the Thomson regime, we have $L_c/L_s = U^\prime_{\rm syn}/U_B$,
we have the third requirement
\begin{equation}
B\delta^2 R \, = \, \left({ 2 L_{\rm syn}^2\over c L_c}\right)^{1/2}
\label{bd2r}
\end{equation}
Equations \ref{bd3}, \ref{bd} and \ref{bd2r} form a closed
system of equations in the unknowns $B$, $\delta$ and $R$.
The solutions are
\begin{equation}
\delta \, = \, \left({ a\over b}\right)^{1/2};~~~~~
B \, = \, { b^{3/2}\over a^{1/2}};~~~~~
R \, = \,  \left({ 2 L_{\rm syn}^2\over c L_c}\right)^{1/2}   { 1\over (ab)^{1/2}}
\end{equation}
where
\begin{equation}
a \, \equiv \,  \left({h \over m_ec^2}\right)^2 \, 
{ \nu_x^3 \over 3.6 \times 10^6}; \qquad
b \, \equiv \,{\nu_x^2 \over 2.8 \times 10^6 \nu_c}
\label{ab}
\end{equation}
For illustration, consider the case $\nu_s=1$ keV, $\nu_c=0.5$ TeV, 
$\nu_x=$ 10 keV, and equal synchrotron and self Compton luminosities,
$L_{\rm syn}=L_c=10^{45}$ erg s$^{-1}$.  Then we obtain $B=0.14$ G,
$\delta=1200$ and $R=1.2\times 10^{12}$ cm.  With these parameters,
the cooling frequency after one hour (observed time) is of the order
of $5\times 10^{16}$ Hz.  The minimum variability timescale, fixed by
the light crossing time, is $t_{\rm var}\sim R/c\delta \simeq 0.03$ s.
As we can see, the required beaming factor is rather extreme for this
reason we consider this solution very unlikely.

\section{Solution of the kinetic equation}
\label{app_solution}

Analytical solution of the kinetic equation used in subsection \ref{sub_rad-cool} is
given by following formula
\begin{eqnarray}
N^*_e(\gamma, t) & = & \left[k_1 \{ S_1(\gamma, t) / \gamma \}^{n_1} S_2(\gamma, t) \left(1 + 
                        \frac{\gamma}{S_1(\gamma,t) \gamma^0_{\rm brk}} \right)^{n_1-n_2} \right]
                     , \label{row_ewe_ad_ak_ch_rozw}\\[0.2cm]
S_1(\gamma, t) & = & [I_1(t) - \gamma I_2(t)], \nonumber\\
S_2(\gamma, t) & = & \exp \Bigg[- \int^{t}_{t_0} \Big\{\gamma C_{\rm adia}(t') I_2(t)
                 - \gamma C_{\rm adia}(t') I_2(t') \nonumber \\
               & - & C_{\rm adia}(t') I_1(t) - 2 \gamma I_1 (t') C_{\rm cool}(t') \Big\} \nonumber\\
               & \Big / & \Big\{ \gamma I_2 (t') - \gamma I_2 (t) + I_1(t) \Big\} \mathrm{d} t' 
               \Bigg], \nonumber\\
I_1(x)         & = & \exp \left[- \int_{t_0}^{x} C_{\rm adia}(y) \mathrm{d}y\right],~~~~
I_2(x)          =  \int_{t_0}^{x} C_{\rm cool}(y) I_1(y) \mathrm{d}y,\nonumber
\end{eqnarray}
This solution describes evolution of the electron energy spectrum only for the case where the 
initial spectrum is approximated by the continuous broken power law function 
(Eq. \ref{equ_ini_elec_spec_rad_cool}).

\end{document}